# Reynolds-number evolution of wall-pressure statistics beneath canonical turbulent boundary layers

**Rahul Deshpande[1]*,† Balaraman Panneerselvam[2]*, Vijaya R. R. Gudla[2], Joe Klewicki[2] and Ivan Marusic[2,3]**

[1]School of Engineering, RMIT University, Melbourne, VIC 3000, Australia
[2]Department of Mechanical Engineering, University of Melbourne, VIC 3010, Australia
[3]Faculty of Mechanical Engineering and Naval Architecture, University of Zagreb, Croatia

This study investigates the Reynolds-number evolution of wall-pressure statistics beneath zero-pressure-gradient turbulent boundary layers, and links their logarithmic variation to the increasingly energetic large-scale motions in the logarithmic (inertial) region. The wall-pressure skewness is found to become more negative with increasing Reynolds number, owing to increasing contributions from large-scale wall-pressure fluctuations (that are negatively skewed) and their nonlinear interaction with the statistically invariant inner-scale fluctuations (that are positively skewed). The analysis draws on new, well-resolved simultaneous measurements of wall pressure and streamwise velocity spanning $5000 \lesssim Re_\tau \lesssim 11,300$ in the Melbourne boundary-layer tunnel, atmospheric surface-layer measurements at $Re_\tau = O(10^6)$ and a published simulation dataset at $Re_\tau = O(10^3)$. Particular attention is paid to the principal experimental limitations affecting wall-pressure statistics: spatial resolution, Helmholtz resonance, facility noise and statistical convergence. Helmholtz resonance is shown to contaminate inner-scale wall-pressure contributions even after conventional corrections, and reliable estimation of skewness is found to require acquisition durations of $O(10^5)$ eddy-turnover times or longer. The inner-scaled wall-pressure spectrum is Reynolds-number invariant over the small-scale regime, in contrast to turbulent channel and pipe flows, whereas at intermediate and large scales it grows substantially with Reynolds number, consistent with these internal flows. Linear and quadratic velocity–wall-pressure coherence analyses link these intermediate- and large-scale contributions to two dynamically distinct classes of coherent structure: the self-similar attached-eddy hierarchy and turbulent superstructures, respectively. Together, these analyses establish that the logarithmic variation of wall-pressure variance and skewness with $Re_\tau$ reflects, respectively, the increasingly strong linear superposition and nonlinear modulation imposed by these logarithmic-region motions. This logarithmic scaling of skewness provides an empirical and physical basis for stochastic-estimation approaches and predictive wall-pressure models at high Reynolds numbers.

* both these authors contributed equally and are joint first-authors
† Email address for correspondence: raadeshpande@gmail.com

Abstract must not spill onto p.2

## 1. Introduction and context

Wall-pressure ($p_w$) fluctuations, generated by the unsteady passage of turbulent motions above a surface, are the primary source of flow-induced structural vibration and the subsequent generation of aerodynamic noise in wall-bounded flows (Willmarth 1975). Although the various canonical wall-bounded flows share similar near-wall turbulence dynamics, they differ in the organisation of their outer-layer motions owing to geometric confinement effects (Monty *et al.* 2009; Lee & Sung 2013). These differences are expected to influence the turbulent flow mechanisms responsible for generating wall-pressure fluctuations and, consequently, the Reynolds-number dependence of the $p_w$-spectrum, a key input for predictive models of turbulent flow-induced noise (Corcos 1964). Establishing the extent to which the wall-pressure statistics are universal across canonical wall flows, and identifying where they differ, therefore remains an important problem in both wall-turbulence research and engineering applications.

Accurate characterisation of these effects requires well-resolved measurements capable of capturing both the time-periodic as well as rare, high-amplitude wall-pressure events occurring over a broad range of temporal frequencies. These are commonly quantified using the broadband $p_w$-spectrum and higher-order statistics (e.g. skewness), respectively. While direct numerical simulations are now available for fully developed turbulent channel and pipe flows across approximately a decade in friction Reynolds number ($O(10^3) \lesssim Re_\tau \lesssim O(10^4)$; Panton *et al.* 2017; Yu *et al.* 2022; Pirozzoli & Wei 2025), comparably well-resolved simulations and/or measurements for the zero-pressure-gradient (ZPG) turbulent boundary layer (TBL) remain relatively scarce (Tsuji *et al.* 2007, 2012; Sillero *et al.* 2014). The present study reports new, well-resolved measurements of $p_w$ beneath ZPG TBLs across $O(10^3) \lesssim Re_\tau \lesssim O(10^4)$ to investigate the Reynolds-number dependence of the $p_w$-spectrum and its skewness, while also relating these statistics to the wall-pressure–velocity correlations within the overlying turbulent boundary layer. Here, $Re_\tau = U_\tau \delta/\nu$, where $U_\tau$ is the mean friction velocity, $\delta$ is the boundary-layer thickness, and $\nu$ is the kinematic viscosity. Throughout this manuscript, $x$, $y$, and $z$ denote the streamwise, spanwise, and wall-normal directions, respectively, with $u$, $v$, and $w$ representing the corresponding velocity fluctuations. Capitalisation or overbar denotes time-averaging, while $\langle \cdot \rangle$ represents conditional averaging.

### 1.1. *Wall-pressure spectrum and its coupling with the overlying flow*

Wall-pressure fluctuations satisfy the pressure Poisson equation and therefore depend on turbulent motions throughout the semi-infinite region above the wall (Willmarth 1975; Tsuji *et al.* 2007). Therefore, pressure-generating motions located beyond the channel or pipe centreline contribute to the measured $p_w$ signatures in internal flows (Abe *et al.* 2005), whereas no analogous contribution is expected in external flows under negligible freestream turbulence. This fundamental distinction motivates a comparative investigation of wall-pressure statistics across canonical internal and external wall flows. To this end, figures 1(a–c) compile premultiplied $p_w$ spectra from published numerical datasets of fully developed turbulent channel (Panton *et al.* 2017) and pipe flows (Yu *et al.* 2022), together with numerical ZPG TBL datasets (Eitel-Amor *et al.* 2014; Deshpande *et al.* 2025). Also included in figure 1(c) are experimental ZPG TBL spectra from Fritsch *et al.* (2022) spanning $4000 \lesssim Re_\tau \lesssim 8400$. Here, $T^+ = U_\tau^2/(f\nu)$ denotes the viscous-scaled time scale, where $f$ is the frequency and the superscript '+' indicates viscous (inner) scaling. Across all three canonical flows, the $p_w$-spectra exhibit a clear Reynolds-number growth over the intermediate and large scales, consistent with previous observations (Tsuji *et al.* 2007; Panton *et al.* 2017; Baars *et al.* 2024; Dacome *et al.* 2025; Pirozzoli & Wei 2025; Deshpande *et al.* 2025). In contrast, notable differences emerge at the smallest viscous scales ($T^+ < O(10^2)$).

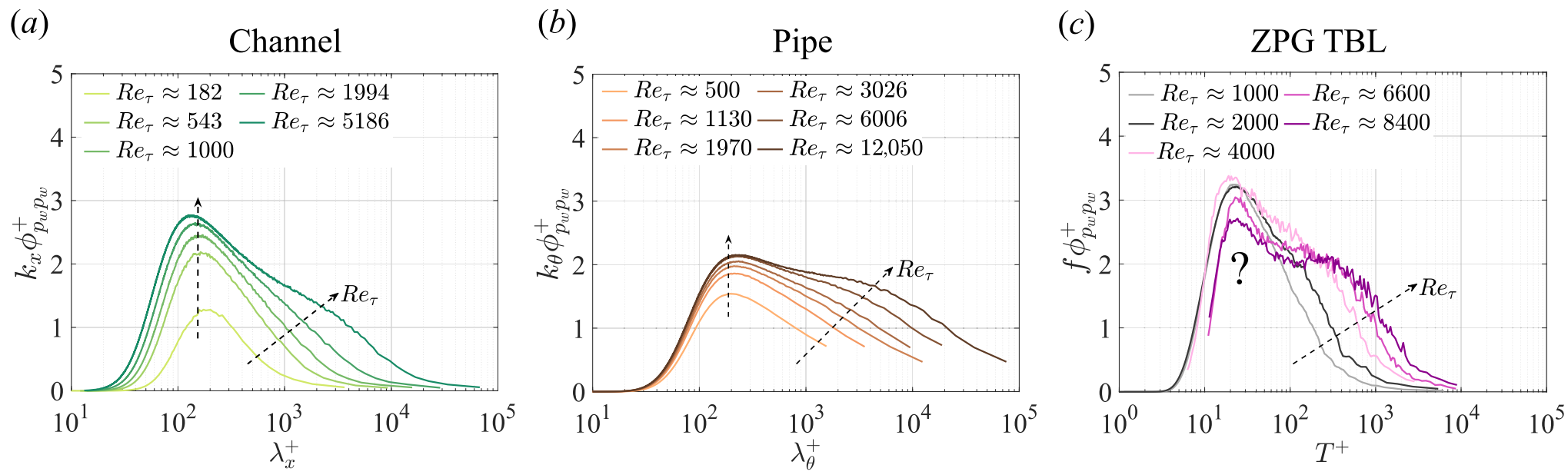

Figure 1: Pre-multiplied frequency/wavenumber spectra of $p_w$ at various $Re_\tau$ for (a) fully developed turbulent channel flows (Panton *et al.* 2017), (b) pipe flows (Yu *et al.* 2022), and (c) ZPG TBLs. In (c), $Re_\tau \approx 1000$ and 2000 are from Eitel-Amor *et al.* (2014), while $4000 \lesssim Re_\tau \lesssim 8400$ are from the experiments of Fritsch *et al.* (2022). $\lambda^+_x = 2\pi/k^+_x$ and $\lambda^+_\theta = 2\pi/k^+_\theta$ are viscous-scaled streamwise and azimuthal wavelengths respectively, with $k_x$ and $k_\theta$ the corresponding wavenumbers used for premultiplication.

The channel and pipe spectra continue to exhibit Reynolds-number growth, although the growth rate progressively weakens with increasing $Re_\tau$. By comparison, the well-resolved ZPG TBL simulations show negligible variation across $1000 \lesssim Re_\tau \lesssim 2000$, whereas the experimentally acquired spectra decrease in magnitude with increasing $Re_\tau$ (the latter most likely reflecting limitations associated with spatial resolution and/or Helmholtz-resonance correction that will be discussed later).

Whether these contrasting trends arise from the absence of pressure-generating motions beyond the channel/pipe centreline, or from differences in the large-scale organisation of turbulence across internal and external flows, cannot be determined from the $p_w$ spectra alone. Resolving this question instead requires simultaneous measurements of wall-pressure and velocity fluctuations (Naka *et al.* 2015) to identify the turbulent motions statistically coupled to $p_w$, and hence associated with the observed Reynolds-number evolution of its statistics. Such scale-dependent coupling between $p_w$ and the overlying turbulent flow has previously been examined over approximately a decade in $Re_\tau$ in turbulent channel and pipe flows by Baars and coworkers (Baars *et al.* 2024; Dacome *et al.* 2025), using linear and quadratic coherence spectra. Importantly, however, this statistical coupling does not by itself identify the physical sources of wall pressure, which is a spatially nonlocal response to the full source field of the pressure Poisson equation. These studies by Baars and coworkers found both coherence metrics to collapse within the intermediate-scale range when scaled with distance from the wall, along $TU(z)/z \approx 14$, supporting a statistical coupling between wall-pressure fluctuations and the geometrically self-similar attached-eddy hierarchy (*i.e.*, motions scaling with distance from the wall). Here, $U(z)$ denotes the mean streamwise velocity at wall-normal location $z$, representing the average convection velocity of the turbulent motions. The same self-similar scaling was subsequently observed in relatively low-$Re_\tau$ ($\approx 2000$) ZPG TBLs by Baars *et al.* (2024), suggesting that the large-scale wall-pressure−velocity coupling is statistically similar across canonical wall-bounded flows (Dacome *et al.* 2025). Beyond the linear coupling, Baars and co-workers (Baars *et al.* 2024; Dacome *et al.* 2025) interpreted the quadratic coherence as a statistical measure of the nonlinear coupling between large-scale $u$-fluctuations and $p_w$ through an amplitude-modulation-type mechanism (Thomas & Bull 1983; Luhar *et al.* 2014; Tsuji *et al.* 2016). This analysis was supported by the close agreement between the coherence spectrum of $u$ with $p_w^2$ and that of $u$ with the Hilbert envelope of $p_w$ (Dacome *et al.* 2025), the latter being a widely adopted measure of quantifying amplitude modulation in wall turbulence (Mathis *et al.* 2009; Tsuji *et al.* 2016).

Despite these advances, simultaneous wall-pressure–velocity measurements in ZPG TBLs remain severely limited. Previous experimental studies relating turbulent structures to wall-pressure fluctuations through linear coherence analyses have largely been restricted to relatively low-to-moderate Reynolds numbers ($Re_\tau \lesssim 3300$; Gibeau & Ghaemi 2021; Baars *et al.* 2024; Deshpande *et al.* 2025; Butt *et al.* 2026). To the authors' knowledge, Baars *et al.* (2024) remains the only study to investigate amplitude modulation of wall-pressure in a ZPG TBL using quadratic coherence, and this was limited to a single Reynolds number ($Re_\tau \approx 2000$). Moreover, the recent high-$Re_\tau$ pipe-flow study of Dacome *et al.* (2025) considered simultaneous $p_w$–$u$ measurements at only two wall-normal locations, precluding a comprehensive assessment of the wall-normal regions and turbulent scales associated with the Reynolds-number variation of the quadratic coherence. Addressing these limitations in the context of a high-$Re_\tau$ ZPG TBL forms a central objective of the present study.

### 1.2. *The skewness of wall-pressure fluctuations*

While the preceding studies have considerably advanced our understanding of the second-order statistics of wall-pressure fluctuations, substantially less is known about their higher-order behaviour. In particular, the skewness of $p_w$, $\mathcal{S}_{p_w}$, provides a statistical measure of the asymmetry of the wall-pressure distribution and is therefore sensitive to intermittent, high-amplitude events that cannot be inferred from the $p_w$-spectrum alone. In a ZPG TBL at $Re_\tau \approx 770$, for example, Ghaemi & Scarano (2013) reported that these extreme wall-pressure events reach amplitudes of 2–3 times the root-mean-square of $p_w$ and occur with a probability of only $\approx 13\%$, yet contribute nearly 60% of $\overline{p_w^2}$. These high-amplitude events also cause the probability density function of $p_w$ to depart markedly from Gaussian, making them directly relevant to statistical models of wall pressure (Tsuji *et al.* 2007; Luhar *et al.* 2014; Gibeau & Ghaemi 2021). Identifying the turbulent scales associated with these extreme events is therefore essential for understanding and ultimately controlling flow-induced structural vibration and aerodynamic noise.

Existing measurements of $\mathcal{S}_{p_w}$, however, have largely been restricted to low-to-moderate Reynolds numbers ($Re_\tau \lesssim 4200$; Schewe 1983; Andreopoulos & Agui 1996; Gravante *et al.* 1998; Tsuji *et al.* 2007; Gibeau & Ghaemi 2021; Baars *et al.* 2024), with relatively little discussion on the statistical convergence and measurement accuracy required for reliable estimation of skewness. Consequently, it remains unknown whether the wall-pressure distribution remains approximately symmetric as Reynolds number increases, or whether increasingly intermittent pressure events lead to a systematic Reynolds-number dependence of $\mathcal{S}_{p_w}$. Resolving this question is also important for stochastic-estimation-based predictive frameworks, which commonly assume the probability density function of wall-pressure fluctuations to be symmetric (Naguib *et al.* 2001; Baars *et al.* 2024). Extending these approaches to high Reynolds numbers therefore first requires establishing the Reynolds-number dependence of $\mathcal{S}_{p_w}$ and identifying the turbulent motions correlated with its evolution. This forms a second central objective of the present study, in which we exploit the relationship between skewness and the amplitude-modulation coefficient, originally established for the streamwise velocity by Mathis *et al.* (2009, 2011), to interpret the mechanism underlying the Reynolds-number dependence of $\mathcal{S}_{p_w}$.

### 1.3. *Factors influencing the accuracy of wall-pressure measurements*

Answering the above questions critically depends on obtaining well-resolved wall-pressure measurements. Although the pioneering investigations documenting the spectral characteristics of $p_w$ date back more than half a century (Willmarth & Wooldridge 1962; Willmarth 1975), the pressure transducers available at the time were often large relative to the characteristic turbulent length scales. This led to attenuation of the small-scale wall-pressure fluctuations together with artificial amplification caused by sensor-induced resonance. Subsequent work by Gravante *et al.* (1998) demonstrated that a viscous-scaled sensor diameter of $d_p^+ < 18$ is required to avoid spectral attenuation while ensuring convergence of higher-order statistics. Here, $d_p$ denotes the diameter of the pinhole through which the microphone is exposed to the flow. The importance of this requirement is evident from the measurements of Fritsch *et al.* (2022) shown in figure 1(c), where $d_p^+ \approx 39$, 67, and 68 for $Re_\tau \approx 4000$, 6600, and 8400, respectively.

Besides spatial resolution, accurate wall-pressure measurements require careful treatment of both facility-induced acoustic contamination and Helmholtz resonance associated with the microphone mounting cavity (Tsuji *et al.* 2007; Gibeau & Ghaemi 2021; Baars *et al.* 2024; Dacome *et al.* 2025). Acoustic contamination primarily affects the low-frequency portion of the spectrum (high $T^+$), whereas Helmholtz resonance distorts the high-frequency range (low $T^+$). The latter becomes increasingly restrictive when high Reynolds numbers are achieved by increasing the freestream velocity, since the accompanying reduction in viscous length scale broadens the energetic wall-pressure spectrum towards higher frequencies. As a result, the Helmholtz resonance bandwidth increasingly overlaps with the energetic high-frequency/small-scale portion of the spectrum, influencing both the magnitude and phase of the Fourier modes in a manner that cannot be fully corrected for (see Dacome *et al.* (2025) and Appendix A for further discussion). Although the acoustic contamination can be substantially reduced through measurements in anechoic facilities (Jiang *et al.* 2025), implementing such facilities for high-$Re_\tau$ turbulent boundary-layer experiments remains challenging because of the large-scale wind tunnels required. Collectively, these limitations have hindered the acquisition of accurate, well-resolved wall-pressure measurements in high-$Re_\tau$ turbulent boundary layers, thereby limiting our understanding of the Reynolds-number dependence of both the second- and higher-order wall-pressure statistics.

### 1.4. *Present contributions*

The present study is designed to overcome the experimental limitations outlined above and thereby obtain well-resolved wall-pressure measurements beneath high-$Re_\tau$ ZPG turbulent boundary layers. The experiments are conducted in the large boundary-layer wind tunnel facility at the University of Melbourne, which follows the 'big and slow' philosophy of achieving high Reynolds numbers through the development of a thick turbulent boundary layer while maintaining comparatively large viscous length scales (Marusic *et al.* 2015).

This approach simultaneously relaxes spatial-resolution requirements and delays the onset of Helmholtz-resonance limitations, enabling accurate wall-pressure measurements across $10^3 \lesssim Re_\tau \lesssim 10^4$. Facility-induced acoustic contamination is estimated and removed using a Wiener-filtering procedure following Hayes (1996). The microphone-mounting cavity is designed so that the Helmholtz resonance lies outside the energetic portion of the wall-pressure spectrum without compromising spatial resolution, and sufficiently long acquisition times are used to ensure statistical convergence of $\mathcal{S}_{p_w}$ across the full Reynolds-number range. The wall-pressure measurements are acquired simultaneously with microscale hot-wire measurements of $u$ spanning the full boundary-layer thickness, enabling identification of the turbulent motions associated with the Reynolds-number evolution of both the second- and third-order wall-pressure statistics.

The laboratory measurements are complemented by published well-resolved large-eddy simulation datasets of ZPG TBLs (Eitel-Amor *et al.* 2014) at $Re_\tau \sim 1000$–2000 to address the following outstanding questions raised in §1: (i) Does the small-scale region of the $p_w$-spectrum exhibit inner scaling, or does it grow with Reynolds number analogous to internal flows? (ii) How does $\mathcal{S}_{p_w}$ evolve with increasing $Re_\tau$, and how do the different regimes of the $p_w$-spectrum contribute to its variation? (iii) Which turbulent scales and wall-normal regions are associated with the Reynolds-number dependence of $\mathcal{S}_{p_w}$? In addition to the well-resolved data, the present study also considers atmospheric surface-layer measurements acquired over the Utah salt flats at $Re_\tau \sim 10^6$ (Klewicki *et al.* 2008) to examine whether the trends observed in laboratory flows persist towards asymptotically high Reynolds numbers.

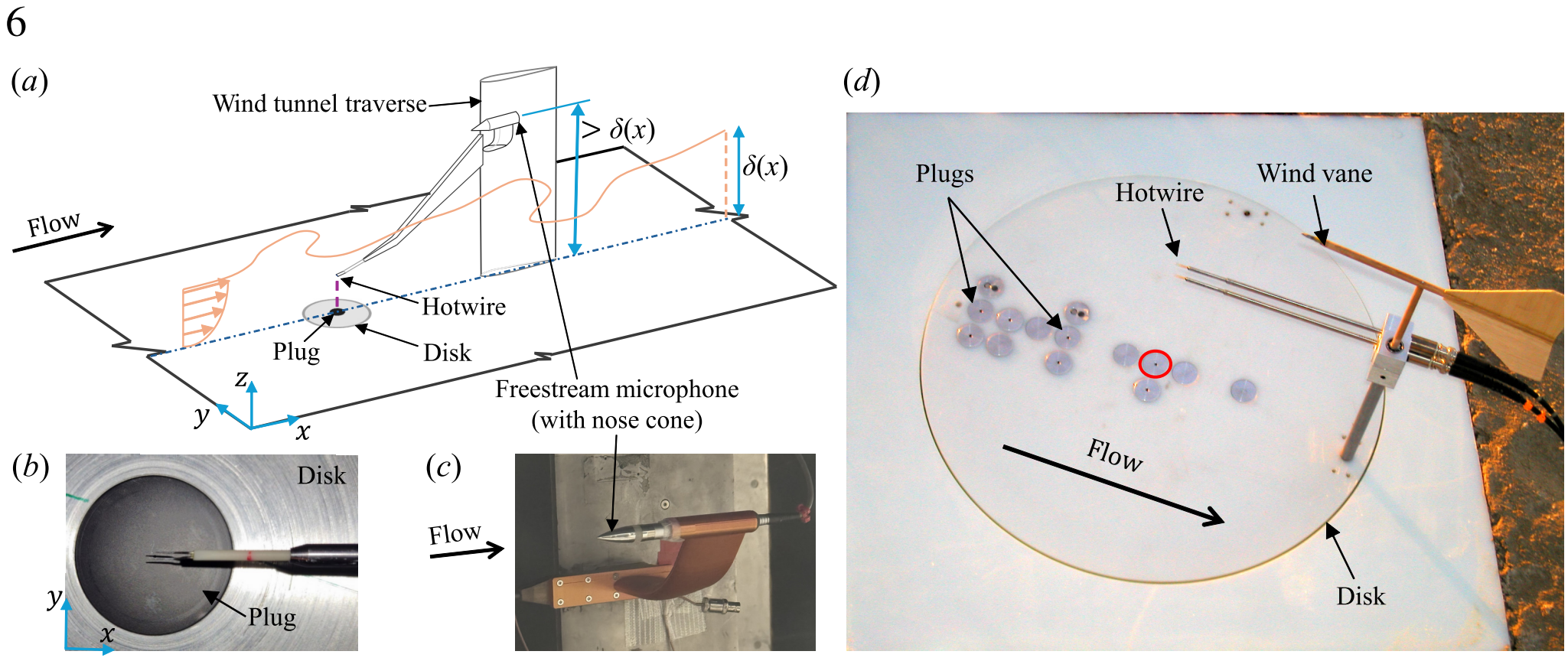

Figure 2: (a-c) Experimental setup in the HRNBLWT facility: (a) a simplified schematic of the setup at $x$ = 13.2 m. (b) Simultaneous hotwire and wall-pressure measurements, with the hotwire positioned vertically above the pinhole of the flush-mounted microphone plug. (c) Freestream microphone fitted with a nosecone and used for facility-noise estimation, procedure for which is described in the supplementary material. (d) Photograph of the SLTEST atmospheric-surface-layer experiment (Klewicki *et al.* 2008), showing multiple wall-mounted microphones and overlying hotwire sensors. The microphone marked by the red ring was vertically below the hotwires, and was used for the $u$–$p_w$ correlation analysis in §3.3.

## 2. Experimental setup, methodologies and previously published datasets

### 2.1. *Wind tunnel experiments across* $O(10^3) \lesssim Re_\tau \lesssim O(10^4)$

Laboratory experiments were conducted in the High Reynolds Number Boundary Layer Wind Tunnel (HRNBLWT) facility at the University of Melbourne, one of the few lab-based facilities capable of generating high-$Re_\tau$ ZPG TBLs while maintaining sufficiently large viscous length scales for obtaining well-resolved measurements. It has a test-section length of 27 m and a cross-sectional area of 1.89×0.92 $m^2$ (width × height). The freestream turbulence intensity is approximately 0.05% at the test-section entrance that increases modestly to 0.15–0.2% over the range 13 m $\lesssim x \lesssim$ 18 m, where $x$ denotes the streamwise distance from the boundary-layer trip. A statistically two-dimensional ZPG TBL is maintained through adjustment of roof-mounted bleed slots, spanning the tunnel width. Further details regarding the facility and its flow characteristics are provided by Marusic *et al.* (2015).

Measurements were conducted at $x$ = 13.2 m and $x$ = 18.2 m, with the corresponding experimental parameters summarised in table 1. Figure 2(a) shows the measurement configuration at $x$ = 13.2 m, where a hotwire sensor was positioned directly above a wall-mounted microphone to obtain simultaneous measurement of $u$ and $p_w$. Both sensors were located at the spanwise centreline of the tunnel and at the same streamwise position. A second station was installed at $x$ = 18.2 m exclusively for wall-pressure measurements, for an independent assessment of spatial-resolution effects and influence of the overlying hotwire sensor (if any) on $p_w$-statistics.

Wall-pressure fluctuations at both stations were measured using a 1/2-inch infrasound microphone (Brüel & Kjær 4193-L-004), the same model employed in the atmospheric surface-layer measurements of Klewicki *et al.* (2008). The microphone has a nominal

(A) Wind tunnel experiments across $O(10^3) \lesssim Re_\tau \lesssim O(10^4)$:

| $x$ (m) | HW | $U_\infty$ (ms$^{-1}$) | $\nu/U_\tau$ ($\mu$m) | $\delta$ (m) | $Re_\tau$ | Mic. $d_p^+$ | HW $l^+$ | HW measurement locations | $TU_\infty/\delta$ | $\Delta t^+$ | Plotting style |
|---|---|---|---|---|---|---|---|---|---|---|---|
| 13.2 | ✓ | 8.4 | 56 | 0.276 | 5000 | 9 | 9 | #40 across: $7 \lesssim z^+ \lesssim \delta^+$ | 19,200 | 0.10 | ♦ — |
| 13.2 | ✓ | 15.0 | 32 | 0.271 | 8300 | 15 | 15 | #40 across: $11 \lesssim z^+ \lesssim \delta^+$ | 18,300 | 0.30 | — |
| 18.2 | × | 10.7 | 43 | 0.346 | 8000 | 12 | - | - | 160,800 | 0.17 | ♦ — |
| 13.2 | ✓ | 21.1 | 23 | 0.261 | 11,300 | 22 | 22 | #40 across: $14 \lesssim z^+ \lesssim \delta^+$ | 18,600 | 0.58 | — |
| 18.2 | × | 14.8 | 31 | 0.338 | 10,700 | 16 | - | - | 157,600 | 0.31 | ♦ — |
| 18.2 | × | 22.2 | 21 | 0.323 | 14,700 | 23 | - | - | 159,300 | 0.65 | — |

(B) Experiments in the atmospheric surface layer at $Re_\tau \sim O(10^6)$:

| HW | $U_\tau$ (ms$^{-1}$) | $\nu/U_\tau$ ($\mu$m) | Mic. $d_p^+$ | HW $l^+$ | HW measurement locations | $TU_\infty/\delta$ | $\Delta t^+$ | Plotting style |
|---|---|---|---|---|---|---|---|---|
| V-type | 0.136 | 132 | 19 | 8 | $z^+ \approx 101$ | 7.0 | 0.21 | * — |
| ×-type | 0.254 | 71 | 35 | 14 | $z^+ \approx 1546$ | 7.9 | 0.72 | * — |
| single | 0.191 | 94 | 27 | 11 | $z^+ \approx 3181$ | 13.6 | 0.41 | * — |

Table 1: Summary of the flow conditions and measurement parameters for the two experimental datasets considered in the present study. Relevant terminology has been introduced/defined in §1 and §2.

sensitivity of 2 mV Pa$^{-1}$ and a frequency response spanning 0.1 Hz to 20 kHz. The sensor was mounted inside an Aluminium plug (6016-T6 alloy), flush with the tunnel floor, and exposed to the wall-pressure field through a pinhole of diameter, $d_p = 0.5$ mm. At the upstream station, the pinhole was positioned directly beneath the hotwire sensor (figure 2b). The plug surface was precision-machined and anodised to minimise roughness effects, yielding wall-pressure statistics consistent with previous smooth-wall ZPG TBL studies (Farabee & Casarella 1991; Eitel-Amor *et al.* 2014; Knoop *et al.* 2026; Deshpande *et al.* 2026). Validation of the measurement arrangement is discussed extensively in §3.1.

Spatial resolution is a critical consideration for measuring a well-resolved $p_w$-spectrum in high-$Re_\tau$ TBLs (Schewe 1983). In this study, we utilise data collected at $x = 13.2$ m for $Re_\tau = 5000$ and at $x = 18.2$ m for $Re_\tau = 8000$ and 10,700, for all analyses focused exclusively on wall pressure statistics (table 1). These measurements satisfy the criteria established by Gravante *et al.* (1998), regarding negligible attenuation of the energetic scales for $d_p^+ < 18$ (significant for $d_p^+ \gtrsim 27$). The downstream station at $x = 18.2$ m was particularly valuable in validating these requirements, given that the desired (larger) $Re_\tau$ could be achieved at relatively lower freestream velocities and larger viscous length scales, yielding relatively better spatial resolution than at $x = 13.2$ m. The resulting spectra were nearly identical between the two stations with only minor attenuation observed for $T^+ \lesssim 2$, confirming that spatial-resolution effects remain negligible over the energetically significant portion of the spectrum for $Re_\tau \lesssim 10{,}700$. The $u$–$p_w$ correlation analyses in the log and outer region are largely insensitive to the modest differences in spatial resolution (Baars *et al.* 2024; Dacome *et al.* 2025), deeming the simultaneous $u$–$p_w$ measurements at $x = 13.2$ m adequate for the present analysis.

At the highest measured Reynolds number, $Re_\tau = 14,700$, however, Helmholtz-resonance effects begin to encroach on the energetic portion of the $p_w$ spectrum and contaminate higher-order statistics such as $\mathcal{S}_{p_w}$; these effects cannot be fully corrected for (see supplementary material and appendix A), and this case is therefore excluded from the main analysis.

The geometry of the microphone-plug assembly determines the spectral signature associated with Helmholtz resonance in the measured wall-pressure signal (Gibeau & Ghaemi 2021; Baars *et al.* 2024). The cavity volume was therefore carefully designed to shift the resonance bandwidth well beyond the energetic region of the $p_w$-spectrum ($T^+ \lesssim 6$; figure 1c). This was verified through microphone-characterisation tests conducted in an anechoic enclosure, which identified a resonance frequency of approximately 3650 Hz corresponding to our microphone-plug assembly. These tests yield a transfer function relating the true pressure signal, measured using a microphone without the plug, to the response of the plug-mounted microphone. The transfer function was subsequently used to correct the raw wall-pressure time series acquired during the wind tunnel experiments for any residual influence of Helmholtz resonance. Full details of the microphone-plug assembly, the anechoic test configuration and the correction methodology are provided in the supplementary material for interested readers.

In addition to Helmholtz resonance effects, microphone measurements acquired in non-anechoic wind tunnels are susceptible to facility-induced acoustic contamination arising from the upstream fan, structural vibrations of the test section, and other background noise sources. While this can be addressed directly (in case of subsonic flows) by removal of the acoustic cone from the frequency-wavenumber spectrum of wall-pressure (Deshpande *et al.* 2026; Butt *et al.* 2026), this methodology can only be deployed when a streamwise array of microphones is used in the measurement. In scenarios when only one wall-pressure microphone is available, previous studies (Gibeau & Ghaemi 2021; Baars *et al.* 2024) have used a Wiener noise-cancelling filter (Hayes 1996) that requires simultaneous measurements of the facility-noise field and the wall-pressure signal. Following this approach, a 1/2-inch Brüel & Kjær infrasound microphone (identical to that used for wall pressure acquisition) was mounted in the freestream, as shown in figures 2(a,c). A streamlined nosecone (HBK UA-0386) was fitted to this microphone to minimise aerodynamic disturbances. The facility-noise component estimated from the freestream microphone was then removed from the Helmholtz-corrected wall-pressure signal to obtain the final corrected wall-pressure time series. The complete implementation of the Wiener filtering procedure follows previous studies (Gibeau & Ghaemi 2021; Baars *et al.* 2024) and is also documented in the supplementary material.

Instantaneous streamwise velocity measurements were acquired using a Platinum hotwire sensor of length $l = 0.5$ mm and diameter $d = 2.5\,\mu$m ($l/d = 200$). The sensor was prepared in-house by first soldering a Wollaston wire to a standard Dantec 55P15 probe. The wire was then exposed to diluted nitric acid that removed the silver coating and exposed the platinum sensor within. During experiments, the sensor was operated using the Melbourne University Constant Temperature Anemometer at an overheat ratio of 1.8. Square-wave testing confirmed an approximately second-order frequency response of the measurement system. During the experiments, instantaneous signals from both microphones and the hotwire were acquired simultaneously at a sampling frequency of $f_s = 50$ kHz. This provided a temporal resolution of $\Delta t^+ (= U_\tau^2/\nu f_s) \lesssim 0.65$. The hotwire signal was routed through an 8-pole Butterworth low-pass filter (Frequency Devices Inc.; cut-off frequency = 25 kHz) before being sent to a data acquisition unit (Data Translation DT9836) to minimise aliasing. Alternatively, the signals from the microphone were first fed to a conditioning amplifier (B&K 2690 0S4) and then to the data acquisition unit.

The hotwire voltage was converted to velocity using a calibration performed against freestream Pitot-tube measurements. The Pitot tube and a thermistor were mounted side-by-side on the wind-tunnel traverse to continuously acquire the freestream velocity and temperature, respectively, throughout each experiment (Marusic *et al.* 2015). The traverse position was controlled through a motorised leadscrew mechanism using an encoder feedback, allowing precise positioning of the hotwire throughout the boundary-layer thickness. At $x = 13.2$ m, velocity measurements were acquired at 40 logarithmically spaced wall-normal locations spanning the entire boundary layer, with simultaneous wall-pressure measurements obtained at each location (table 1). The acquisition duration ($T$) at each wall-normal position was selected such that the total eddy-turnover time satisfied $TU_\infty/\delta \gtrsim 18,000$, ensuring convergence of the large-scale spectral statistics. At the downstream station ($x = 18.2$ m), where only wall-pressure measurements were performed, substantially longer acquisition durations were employed ($TU_\infty/\delta \approx 160\,000$) to assess convergence of higher-order statistics, particularly the wall-pressure skewness $\mathcal{S}_{p_w}$ (see Appendix B). The boundary-layer thickness was determined using the methodology proposed recently by Lozier *et al.* (2025), based on the unique behaviour of the streamwise velocity skewness profile near the boundary-layer edge. Given the sensitivity of the normalised wall-pressure spectrum to the friction velocity, oil-film interferometry was used to measure $U_\tau$ directly.

### 2.2. *Experiments in the atmospheric surface layer at $Re_\tau \sim O(10^6)$*

The atmospheric surface-layer datasets considered in this study were acquired during the Surface Layer Turbulence and Environmental Science Test (SLTEST) campaign conducted over the Utah salt flats between 2003 and 2004, a small subset of which has previously been published in Klewicki *et al.* (2008). The present analysis utilises unpublished data comprising simultaneous hotwire and wall-pressure measurements, with the associated experimental parameters summarised in table 1. Although these measurements do not exhibit the same level of statistical convergence as the HRNBLWT dataset, they serve as a valuable reference at Reynolds numbers approaching $Re_\tau \sim O(10^6)$. Figure 2(d) shows the SLTEST measurement configuration, comprising hotwire sensors and an array of wall-mounted microphones arranged in the streamwise and spanwise directions. The hotwires were positioned directly above one of the downstream microphone locations (highlighted by the red ring in figure 2d). While signals from all microphones were utilised to compute ensemble-averaged wall-pressure statistics, only those from the microphone located directly beneath the hotwire sensors were used for the estimation of $u$–$p_w$ correlations owing to their sensitivity to spatial offsets (Deshpande *et al.* 2026; Butt *et al.* 2026).

Although the SLTEST campaign employed a suite of microphone models (Klewicki *et al.* 2008), the present analysis is restricted to data acquired using the same microphone model as that employed in the HRNBLWT experiments (§2.1). Each microphone was mounted within a flush-mounted plug installed in a disk embedded in the salt-flat surface. The isolated location of the SLTEST facility ensured minimal contamination from acoustic noise sources and structural vibrations, except during occasional aircraft flyovers, in which case the affected records were discarded. Furthermore, the Helmholtz resonance bandwidth of the microphone–plug assembly was reported to lie well outside the energetic region of the wall-pressure spectrum (Klewicki *et al.* 2008). Therefore, neither acoustic-noise nor Helmholtz-resonance corrections were required.

Instantaneous streamwise velocity measurements were acquired at three wall-normal locations ($z^+ = 101$, 1546 and 3181) through independent experiments, using V-wire, ×-wire and single-wire probes, respectively. Standard de-trending procedures were applied to remove records affected by transient changes in flow direction and non-stationarity (Klewicki *et al.* 2008; Hutchins *et al.* 2012). In addition, only runs satisfying near-neutral atmospheric conditions, as determined from sonic-anemometer and temperature measurements, were retained for analysis. Depending on atmospheric conditions, the usable duration of individual records varied between approximately 3 to 10 minutes. The hotwire signals were converted to velocity using an in-situ calibration performed with a portable jet-calibration facility, where the hotwire response was referenced against Pitot-tube measurements obtained at the jet centreline. Simultaneous measurements of $u$ and $p_w$ were acquired at a sampling frequency, $f_s = 5000$ Hz. All relevant acquisition parameters and sensor-resolution metrics ($l^+$, $d_p^+$, $\Delta t^+$ and $TU_\infty/\delta$) are summarised in table 1. Notably, the friction velocity was estimated based on the Reynolds shear stresses measured using sonic anemometry in the logarithmic region (Klewicki *et al.* 2008; Hutchins *et al.* 2012), where these stresses nominally equal unity when scaled by $U_\tau^2$. The inevitable variation of the atmospheric flow conditions, however, causes considerable changes in the estimated $U_\tau$ across measurements (table 1), and consequently the friction Reynolds number: $8 \times 10^5 \lesssim Re_\tau \lesssim 1.2 \times 10^6$ (Klewicki *et al.* 2008). For convenience in discussion, however, we will associate the SLTEST data with $Re_\tau \sim O(10^6)$.

Although the spatial and temporal resolutions of the SLTEST measurements are comparable to those achieved in the HRNBLWT experiments (table 1), the SLTEST dataset is limited by the short duration of the usable records ($TU_\infty/\delta < 15$), which stems from unavoidable variations in atmospheric conditions. As a result, the large-scale end of the wall-pressure spectrum and higher-order statistics such as $\mathcal{S}_{p_w}$ remain statistically unconverged (see Appendix B). Combined with uncertainties associated with the transitionally rough nature of the salt-flat surface, these limitations preclude the use of the SLTEST dataset itself for establishing new conclusions. Instead, it is employed primarily to provide qualitative context for the trends observed in the laboratory and numerical datasets, while illustrating the importance of measurement duration and experimental fidelity in wall-pressure studies.

### 2.3. *Published numerical simulations at $Re_\tau = O(10^3)$*

The numerical data considered in this study are based on the high-resolution large-eddy simulation (LES) of a ZPG TBL performed by Eitel-Amor *et al.* (2014), which has been analysed extensively in past wall-pressure studies (Deshpande *et al.* 2025, 2026). The dataset provides a valuable low-$Re_\tau$ reference that complements the present laboratory measurements while avoiding many of the experimental limitations associated with wall-pressure acquisition. The simulation employed a sufficiently large computational domain to permit development of the boundary layer up to $Re_\tau \approx 2000$. The viscous-scaled grid spacings in the streamwise and spanwise directions were $\Delta x^+ = 18$ and $\Delta y^+ = 8$, respectively, corresponding to approximately twice the resolution of a conventional DNS, with the smallest unresolved scales represented through a subgrid-scale model. Time-resolved velocity and wall-pressure signals were sampled at a temporal resolution of $\Delta t^+ (= U_\tau^2/\nu f_s) \lesssim 0.5$ over a total acquisition duration of $TU_\infty/\delta = 243$. The dataset contains synchronised time series of $u$ and $p_w$, sampled over 1152 spanwise grid points, with velocity signals available at multiple wall-normal locations. Hence, although the total acquisition duration is considerably shorter than that of the HRNBLWT experiments, the large number of spanwise grid points permits robust ensemble averaging and yields reasonably converged large-scale spectral statistics (Deshpande *et al.* 2025). Consequently, the LES dataset serves as an important reference for assessing Reynolds-number trends in the wall-pressure spectrum and the scale-dependent coupling between velocity and wall-pressure fluctuations.

## 3. Results and discussion

### 3.1. *Reynolds number variation of wall-pressure variance and spectra*

We first establish the fidelity of the present wall-pressure measurements before examining the Reynolds-number evolution of the wall-pressure spectrum. As discussed in §1 and §2.1, the wall-pressure measurement system and post-processing methodology were specifically designed to obtain broadband spectra with high spatial resolution across $5000 \lesssim Re_\tau \lesssim$ 11, 300, while minimising contamination from Helmholtz resonance and facility noise. Figure 3(a) validates the resulting measurements by comparing the inner-scaled root-mean-square of wall-pressure fluctuations, $\sqrt{\overline{p_w^2}}^+ = \sqrt{\overline{p_w^2}}/(\rho U_\tau^2)$, with well-established empirical relationships for ZPG TBLs (Farabee & Casarella 1991; Panton *et al.* 2017). Also included are statistics from published simulations (Eitel-Amor *et al.* 2014; Sillero *et al.* 2014), wind-tunnel experiments (Fritsch *et al.* 2022), and atmospheric surface-layer measurements (Klewicki *et al.* 2008). The present measurements closely follow the empirical trends of Farabee & Casarella (1991) and Panton *et al.* (2017), confirming the expected logarithmic growth of $\overline{p_w^2}^+$ with increasing $Re_\tau$ and providing confidence in the fidelity of the present measurements.

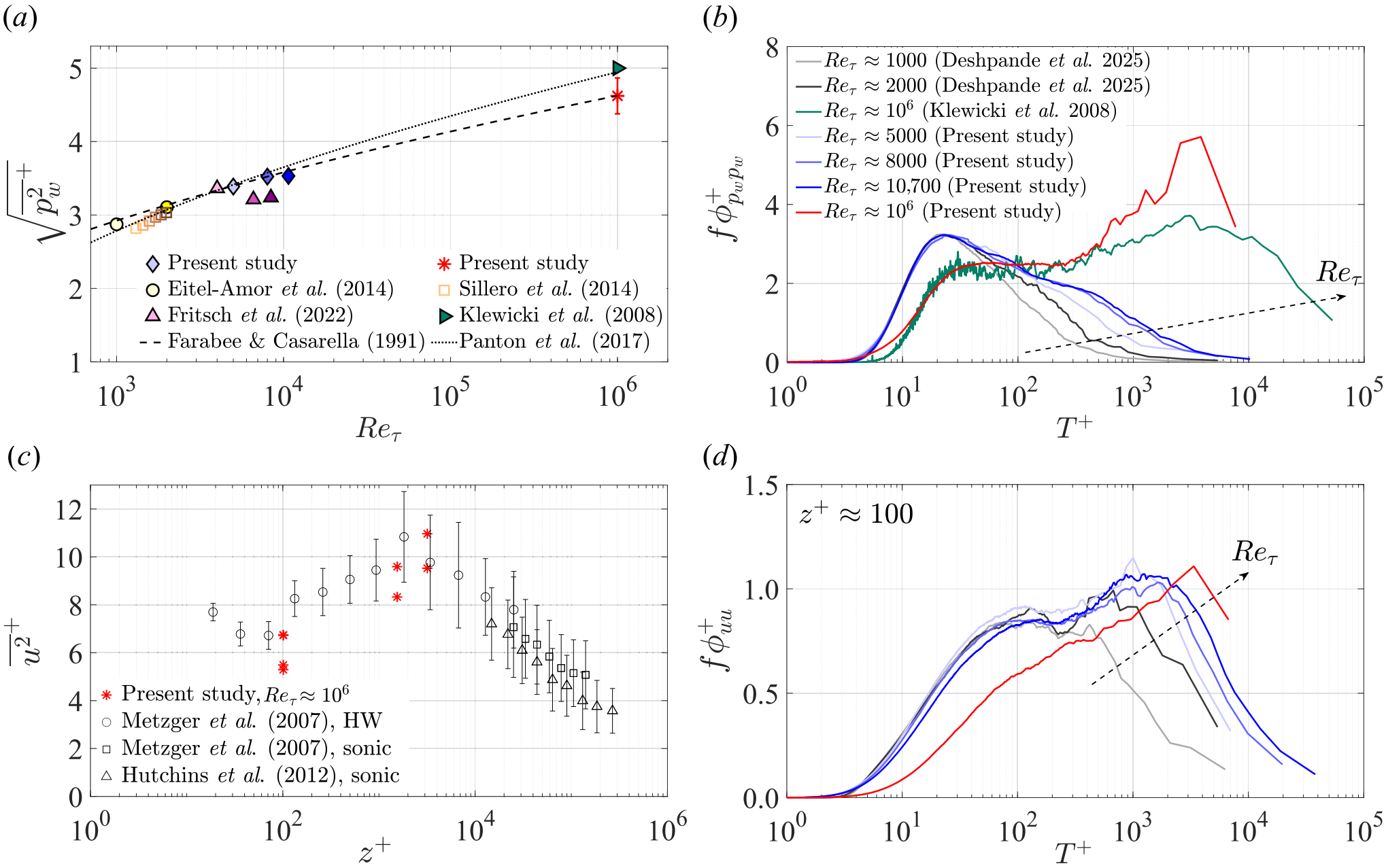

Figure 3: (a) Inner-scaled root-mean-square of $p_w$ from the present experiments compared with various published datasets (Klewicki *et al.* 2008; Sillero *et al.* 2014; Eitel-Amor *et al.* 2014; Fritsch *et al.* 2022). Dashed line represents the empirical fit given by Farabee & Casarella (1991): $\sqrt{\overline{p_w^2}}^+ = \sqrt{6.5 + 1.86\ \ln(Re_\tau/333)}$, while the dotted line represents that given by Panton *et al.* (2017) based on the ZPG TBL data of Schlatter & Örlü (2010): $\sqrt{\overline{p_w^2}}^+ = \sqrt{2.42\ \ln(Re_\tau) - 8.96}$. An increase in colour intensity corresponds to an increase in $Re_\tau$. (b,d) Pre-multiplied frequency spectra of (b) $p_w$ and (d) $u$ at $z^+ \approx 100$ as a function of $T^+$ for various datasets. (c) Inner-scaled variance of $u$ measured by hotwire and sonic anemometry at the SLTEST facility, compared with available literature for $Re_\tau \sim 10^6$ (Metzger *et al.* 2007; Hutchins *et al.* 2012). The error bar indicates $\pm 1$ standard deviation in the ensemble average.

The importance of adequate spatial resolution is highlighted by the measurements of Fritsch *et al.* (2022), which span a Reynolds-number range comparable to that considered here. Their values of $\sqrt{\overline{p_w^2}}^+$ at $Re_\tau \approx 6600$ and 8400 lie systematically below the empirical trends, consistent with the attenuation observed at the small-scale end of the spectrum in figure 1(c). This behaviour is expected because $\overline{p_w^2}^+ = \int_0^\infty \phi^+_{p_w p_w}\, \mathrm{d}T^+$, such that attenuation of the small-scale spectral energy directly reduces the measured wall-pressure variance.

The Reynolds-number trend is further supported by the atmospheric surface-layer measurements at $Re_\tau \approx 10^6$, corresponding to the previously published case of Klewicki *et al.* (2008)

together with the ensemble-averaged estimate obtained from the broader SLTEST dataset considered here (of which the former is a subset). Although these measurements possess adequate spatial resolution, they exhibit greater scatter than the laboratory measurements owing to incomplete convergence of the largest scales, possible surface-roughness effects, and Reynolds-number variations across $8 \times 10^5 \lesssim Re_\tau \lesssim 1.2 \times 10^6$. These uncertainties are reflected by the error bars shown in figure 3(a).

Having established the fidelity of the measurements, we now examine the Reynolds-number evolution of the wall-pressure spectrum. Figure 3(b) compares the premultiplied spectra across $1000 \lesssim Re_\tau \lesssim 10,700$, while also including the SLTEST measurements for completeness. Consistent with the observations from the low-$Re_\tau$ LES datasets in figure 1(c), the present measurements provide compelling evidence that the small-scale end of the $p_w$-spectrum exhibits inner scaling up to approximately $T^+ \lesssim 40$. This behaviour persists across an order-of-magnitude variation in Reynolds number and stands in marked contrast to the channel- and pipe-flow spectra in figures 1(a,b), where the same range of scales exhibits noticeable Reynolds-number growth. To the best of the authors' knowledge, the present dataset provides the first comprehensive experimental verification of inner scaling in the wall-pressure spectrum across such a broad range of $Re_\tau$. The spectra also confirm the robustness of the inner spectral peak at $T^+ \approx 25$ previously reported by Farabee & Casarella (1991) and subsequently observed in pipe flows by Dacome *et al.* (2025).

Beyond the small scales, however, substantial Reynolds-number growth is evident throughout the intermediate- and large-scale ranges, becoming particularly pronounced after including the atmospheric surface-layer measurements. The SLTEST spectrum shown here was obtained by ensemble averaging spectra from multiple microphone records, which are plotted individually in figure 11 of Appendix C. While the ensemble-averaged spectrum compares reasonably well with that reported by Klewicki *et al.* (2008), some differences remain at both the smallest and largest scales, most likely owing to the uncertainties discussed above.

To shed some light on the velocity flow structures associated with this Reynolds-number growth, we consider the corresponding premultiplied streamwise velocity spectra at $z^+ \approx 100$ shown in figure 3(d). This wall-normal location captures contributions from a broad range of inertia-dominated large-scale motions, as demonstrated in previous studies (Baars *et al.* 2024; Deshpande *et al.* 2025). Although the Reynolds-number growth of these velocity motions has been extensively documented (Marusic *et al.* 2015), placing the $u$- and $p_w$-spectra side by side highlights their remarkably similar evolution with increasing $Re_\tau$. The slight offset between the SLTEST and wind-tunnel velocity spectra is likely attributable to the transitionally rough nature of the Utah salt flats, consistent with the observations of Jiménez (2004), who showed that surface roughness preferentially attenuates the small-scale end of the near-wall velocity spectrum.

The similarity between the Reynolds-number evolution of the velocity and wall-pressure spectra naturally motivates the coherence analysis presented in §3.3, where their statistical coupling is examined directly. Since that analysis also incorporates previously unpublished hotwire measurements from the SLTEST campaign, their fidelity is first assessed here in figure 3(c). The inner-scaled streamwise velocity variance, $\overline{u^2}^+ = \overline{u^2}/U_\tau^2$, is compared at various wall-normal locations and for different hotwire configurations (table 1) with the atmospheric measurements of Hutchins *et al.* (2012) and Metzger *et al.* (2007). Agreement is generally within the uncertainty bounds of the earlier studies except at $z^+ \approx 100$, where attenuation of the small-scale velocity spectrum (figure 3d) leads to a slight reduction in the measured variance relative to published statistics. The pressure–velocity coherence analysis in §3.3 therefore considers only the measurements at $z^+ \approx 1500$ and 3000.

Before examining the pressure–velocity coupling, however, we first investigate how the pronounced Reynolds-number growth of the intermediate- and large-scale wall-pressure fluctuations influences the higher-order wall-pressure statistics. In particular, these increasingly energetic fluctuations can modify both the frequency and amplitude of intermittent $p_w$ events, analogous to their influence on wall-shear-stress and near-wall velocity fluctuations (Thomas & Bull 1983; Mathis *et al.* 2009, 2013), thereby altering the statistical symmetry of $p_w$. This motivates the skewness analysis presented next, where the Reynolds-number evolution of moderate- and strongly-energetic wall-pressure events is examined.

### 3.2. *Reynolds number variation of the skewness of wall-pressure fluctuations*

Having established the Reynolds-number evolution of the wall-pressure spectra, we now turn to its influence on the higher-order statistics of wall-pressure fluctuations. In particular, the skewness of $p_w$, $\mathcal{S}_{p_w}$, quantifies the departure of the probability density function (PDF) from Gaussian symmetry and thus provides insight into the prevalence of intermittent moderate- and high-amplitude wall-pressure events (Ghaemi & Scarano 2013). Despite considerable interest in this quantity (Snarski & Lueptow 1995; Andreopoulos & Agui 1996; Lamballais *et al.* 1997; Tsuji *et al.* 2007; Klewicki *et al.* 2008; Gibeau & Ghaemi 2021), no clear Reynolds-number trend has been established to date. Figure 4(a) illustrates this ambiguity: published measurements from canonical wall-bounded flows exhibit substantial scatter (Schewe 1983; Snarski & Lueptow 1995; Klewicki *et al.* 2008; Gibeau & Ghaemi 2021; Baars *et al.* 2024; Dacome *et al.* 2025; Knoop *et al.* 2026), although the majority of well-resolved datasets report $\mathcal{S}_{p_w} < 0$. The present measurements likewise yield negative skewness at all three Reynolds numbers, with $\mathcal{S}_{p_w} = -0.05$ at $Re_\tau = 5000$ and progressively more negative values at higher Reynolds numbers. The variation of $\mathcal{S}_{p_w}$ is well described by a simple logarithmic dependence on $Re_\tau$, obtained via nonlinear least-squares regression of the present well-resolved measurements,

$$\mathcal{S}_{p_w} = 0.0671 \, \ln\left(\frac{2414}{Re_\tau}\right), \tag{3.1}$$

Drawing analogy with the Reynolds-number growth of $\overline{p_w^2}^+$ and other near-wall turbulence statistics (Mathis *et al.* 2013; Pirozzoli & Wei 2025), this logarithmic trend suggests that the evolution of $\mathcal{S}_{p_w}$ is likely associated with the energisation of large-scale inertia-dominated motions. Also, $\mathcal{S}_{p_w}$ can be expected to transition from positive to negative at $Re_\tau = 2414$, indicating the nominal $Re_\tau$ range where the growing large-scale influences 'overpower' the small(inner)-scale contributions. We first establish the robustness of this trend before identifying the turbulent scales and physical mechanisms associated with it.

The substantial scatter among previous measurements suggests that establishing the Reynolds-number dependence of $\mathcal{S}_{p_w}$ requires particular care in both the acquisition and processing of the wall-pressure signal. Based on the available literature and the analyses presented here, the discrepancies amongst the past data plotted in figure 4(a) can be attributed primarily to three factors: insufficient spatial resolution (Schewe 1983), inadequate correction for Helmholtz resonance (Dacome *et al.* 2025), and insufficient statistical convergence. For example, Schewe (1983) showed that increasing the sensor size ($d_p^+$) causes $\mathcal{S}_{p_w}$ to approach zero and, in some cases, become slightly positive. Likewise, the analyses presented in Appendices A and B demonstrate the sensitivity of the estimated skewness to inadequate Helmholtz resonance correction and insufficient sampling duration ($TU_\infty/\delta < 10^5$).

In contrast, the present wind-tunnel measurements across $5000 \lesssim Re_\tau \lesssim 10,700$ satisfy all three requirements. The measurements are well resolved, require only minimal Helmholtz resonance correction, and are acquired over sufficiently long records ($TU_\infty/\delta \approx 1.6 \times 10^5$) to ensure convergence of $\mathcal{S}_{p_w}$. These considerations provide confidence that the Reynolds-number trend represented by equation (3.1) reflects the underlying turbulence rather than measurement artefacts. By comparison, most previously published datasets shown in figure 4(a) were acquired over substantially shorter records ($\lesssim 3 \times 10^4$ turnover times), with the notable exception of Gibeau & Ghaemi (2021), who considered $TU_\infty/\delta \sim 7.5 \times 10^4$. The atmospheric SLTEST estimates are likewise based on considerably shorter records and should therefore be interpreted with appropriate caution. The corresponding uncertainty is reflected by the error bar in figure 4(a), which represents $\pm 1$ standard deviation of the ensemble average and also incorporates uncertainty in the exact flow Reynolds number.

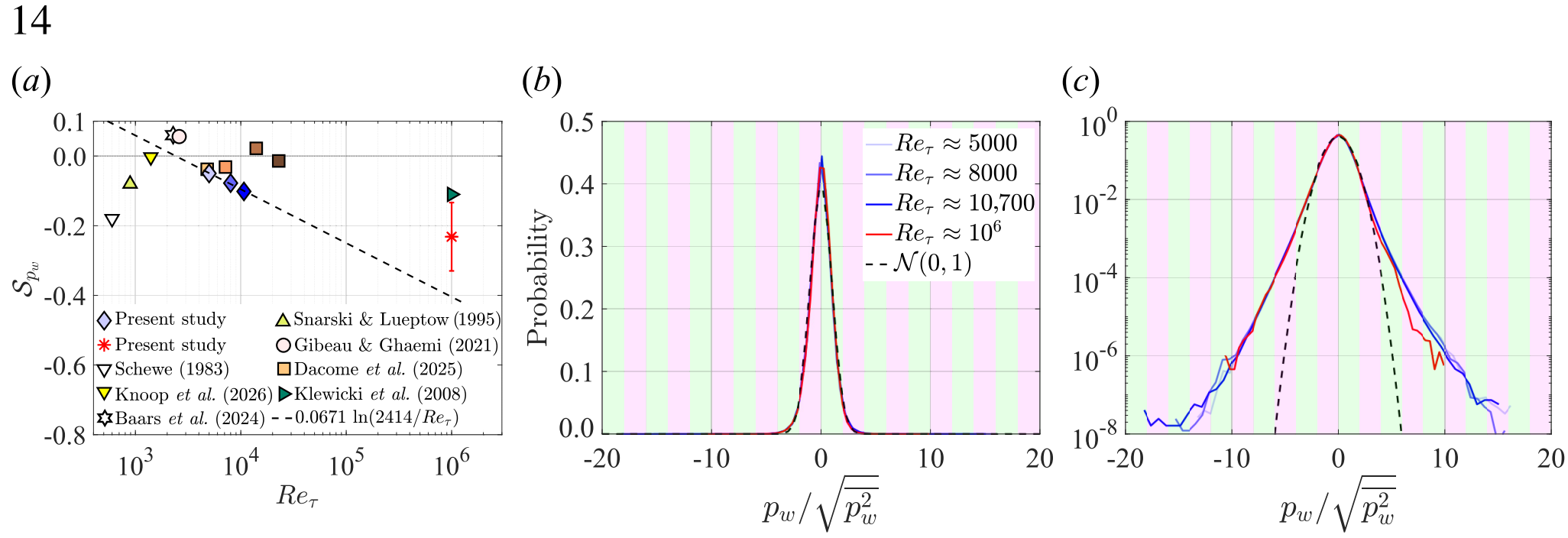

Figure 4: (a) Mean skewness of $p_w$ fluctuations from present and past investigations in canonical wall flows. Only those studies that have reported the length of the $p_w$-time series are considered. The error bar at $Re_\tau \sim 10^6$ indicates $\pm 1$ standard deviation in the ensemble average. (b,c) Probability density functions (PDFs) of $p_w$ fluctuations plotted with (b) linear and (c) logarithmic y-axes against $p_w/\sqrt{\overline{p_w^2}}$. Also plotted is a standard normal distribution given by $\mathcal{N}(0,1)$. Coloured background represents regularly spaced bands of $\Delta p_w = 2\sqrt{\overline{p_w^2}}$, for which we compute the area under the PDF curves presented later in figures 5(a-c).

Next, we investigate which wall-pressure events are responsible for the Reynolds-number variation of $\mathcal{S}_{p_w}$. Since skewness is commonly associated with the tails of the probability density function, one might expect increasingly negative skewness to arise from a greater prevalence of extreme negative pressure events. Figures 4(b,c) test this hypothesis by comparing the PDFs of $p_w$, plotted using linear and logarithmic ordinates, respectively, with $p_w$ normalised by $\sqrt{\overline{p_w^2}}$. Only the statistically converged wind-tunnel datasets, together with the SLTEST measurements and the standard normal distribution, are shown. Surprisingly, the PDFs exhibit no obvious Reynolds-number trend in either representation. In particular, the negative tails do not show a systematic increase in extreme events with increasing $Re_\tau$, unlike the behaviour reported previously for positive tails of the wall-shear stress (Thomas & Bull 1983; Schlatter & Örlü 2010; Mathis *et al.* 2013; Pan & Kwon 2018). This observation suggests that the increasingly negative wall-pressure skewness is unlikely to be driven primarily by rare extreme events. Instead, it points towards a more subtle redistribution of probability associated with moderately strong wall-pressure fluctuations.

To quantify this redistribution, figures 5(a-c) show the integrated area under the PDF over successive intervals of normalised wall-pressure amplitude (*e.g.*, 0–2, 2–4, etc., corresponding to the shaded regions in figures 4b,c) for $Re_\tau = 5000$, 8000 and 10,700, respectively. The positive and negative contributions are evaluated separately, with their integrated areas (marked by circles) respectively connected by solid and dashed lines. For all Reynolds numbers, the probability associated with low-amplitude events ($|p_w| \lesssim 2\sqrt{\overline{p_w^2}}$) remains nearly symmetric. Beyond this threshold, however, a systematic asymmetry emerges. At $Re_\tau = 5000$, negative pressure events occur more frequently than positive events over the range $2 \lesssim |p_w|/\sqrt{\overline{p_w^2}} \lesssim 5$, whereas the most extreme events exhibit the opposite trend (*i.e.*, positive events occur relatively more frequently than negative). As Reynolds number increases to 10, 700, the excess of moderately-strong negative events extends over a substantially broader range: $2 \lesssim |p_w|/\sqrt{\overline{p_w^2}} \lesssim 7$, while the dominance of positive events is restricted only to strong but rare $p_w$ events, resulting in mean negative skewness. Figure 5(d) highlights the former for two representative pressure-amplitude bands, demonstrating a systematic increase in the occurrence of moderately strong negative pressure events with increasing $Re_\tau$. Taken together, these observations demonstrate that the Reynolds-number trend in equation (3.1) is governed primarily by the increasing prevalence of moderately strong negative pressure excursions rather than by positive pressure excursions in the extreme tails of the distribution. This interpretation is also consistent with the observations of Schewe (1983), who similarly reported an excess of negative pressure events beyond approximately $2\sqrt{\overline{p_w^2}}$.

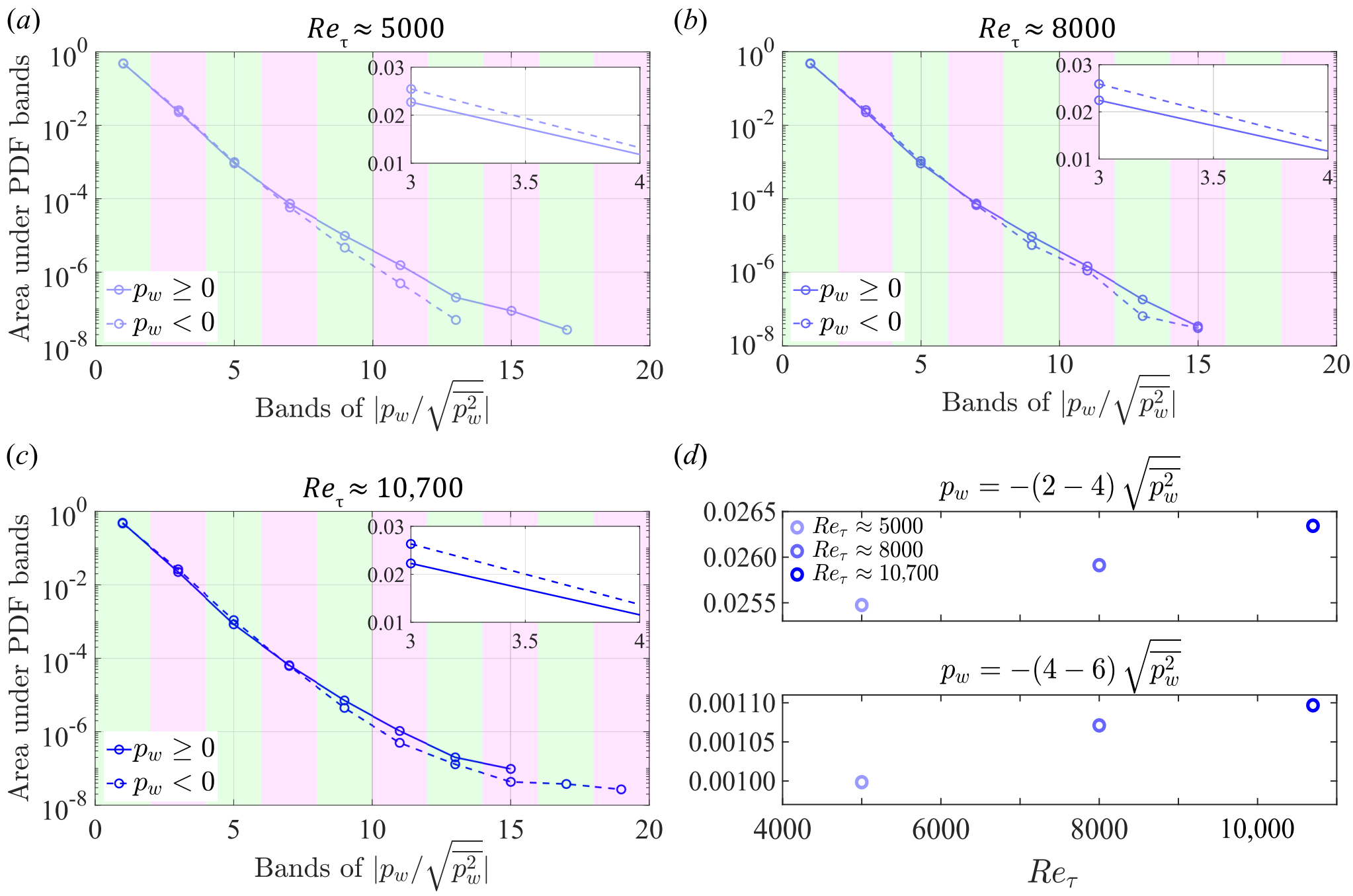

Figure 5: Area under the PDF curves, plotted in figures 4(b,c), computed across regularly-spaced bands of $\Delta p_w = 2\sqrt{\overline{p_w^2}}$ for $Re_\tau \approx$ (a) 5000, (b) 8000 and (c) 10,700. The regularly-spaced bands considered here are indicated by coloured background. Contributions from the positive and negative ends of the PDF($p_w$) are connected via solid and dashed lines, respectively. (d) Area under the PDF for -4 $\leqslant p_w/\sqrt{\overline{p_w^2}} \leqslant$ -2 and -6 $\leqslant p_w/\sqrt{\overline{p_w^2}} \leqslant$ -4 as a function of $Re_\tau$.

Next, we focus on determining which turbulent scales are responsible for the Reynolds-number dependence of $\mathcal{S}_{p_w}$. Motivated by the inner scaling identified in figure 3(b), the wall-pressure signal is decomposed into an inner-scale component ($p_{w,i}$) and a large-scale component ($p_{w,L}$) using a sharp viscous-scaled cutoff timescale of $T_c^+ = 40$. This cutoff corresponds to the upper limit of the Reynolds-number-invariant portion of the wall-pressure spectrum, which is shown again in figure 6(a). The skewness of $p_w$ can then be decomposed

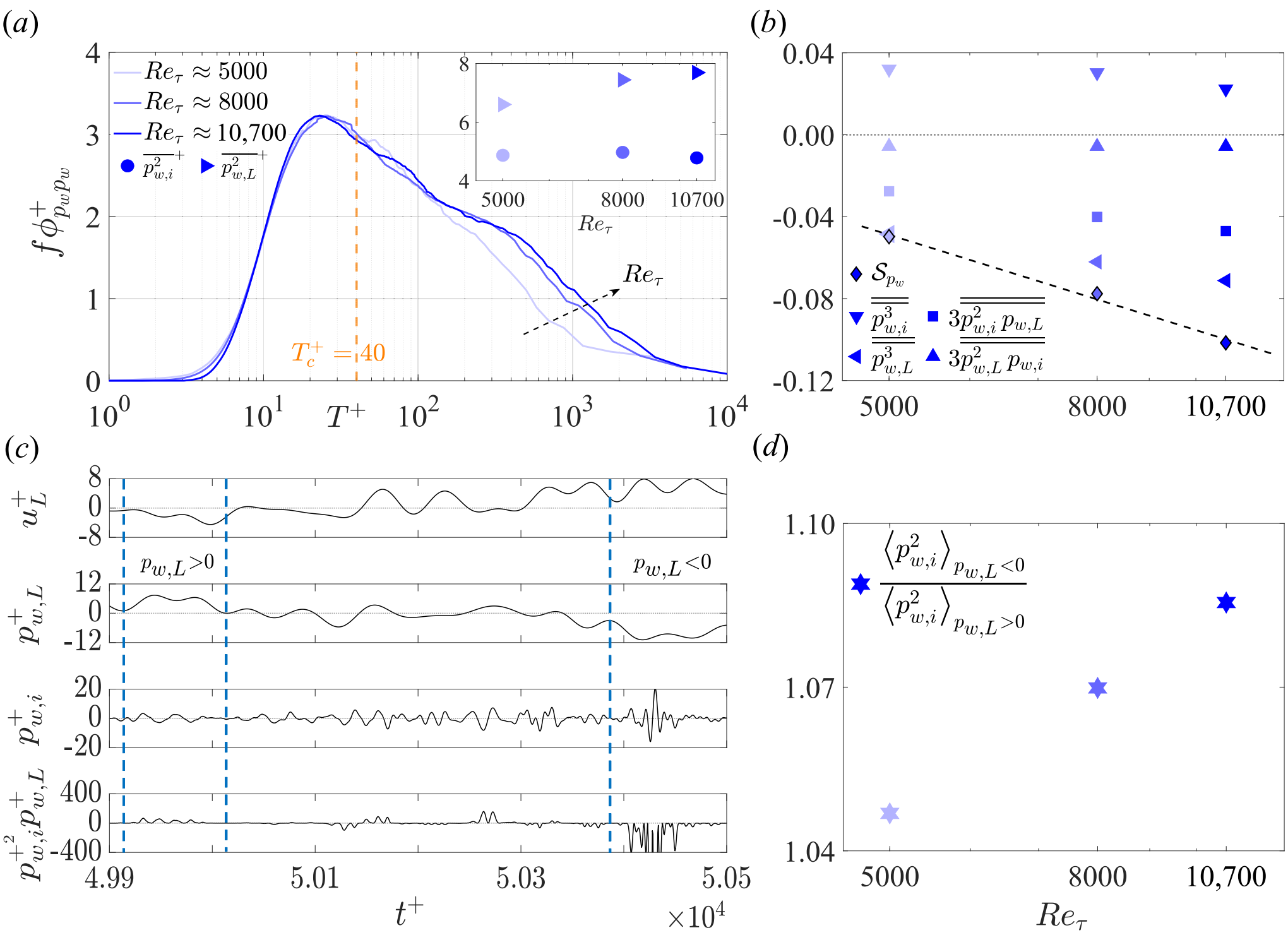

Figure 6: (a) Pre-multiplied frequency spectra of $p_w$ as a function of $T^+$ for HRNBLWT datasets. The inner ($\overline{p^2_{w,i}}^+$) and large-scale ($\overline{p^2_{w,L}}^+$) contributions to the $p_w$-variance are plotted in the inset. (b) Decomposition of the skewness of wall-pressure fluctuations based on $T^+_c = 40$ following (3.2) at various $Re_\tau$. Also plotted is the empirically-obtained relationship in (3.1). (c) Time series of $u^+_L$ (at $z^+ \approx 15$), $p^+_{w,i}$, $p^+_{w,L}$ and $p^{2+}_{w,i}p^+_{w,L}$ at $Re_\tau \approx 11{,}300$, highlighting instants corresponding to $p_{w,L} > 0$ and $p_{w,L} < 0$. (d) Reynolds number variation of the ratio of conditionally-averaged variance of $p_{w,i}$ corresponding to time instants associated with $p_{w,L} < 0$ and $p_{w,L} > 0$.

following Mathis *et al.* (2011) as:

$$\mathcal{S}_{p_w} = \overline{p^3_w} \Big/ \left(\overline{p^2_w}\right)^{3/2} = \overline{\overline{p^3_w}} = \overline{\overline{p^3_{w,i}}} + \overline{\overline{p^3_{w,L}}} + \overline{\overline{3p^2_{w,i}p_{w,L}}} + \overline{\overline{3p^2_{w,L}p_{w,i}}}, \tag{3.2}$$

where $\overline{\overline{(\cdot)}} = (\cdot)/(\overline{p^2_w})^{3/2}$ denotes normalisation by the cube of the root-mean-square pressure. The first two terms represent the intrinsic skewness of the inner- and large-scale components, while the remaining two quantify the nonlinear interaction between them (Mathis *et al.* 2011; Lozier *et al.* 2024). This decomposition enables the Reynolds-number evolution of $\mathcal{S}_{p_w}$ to be directly attributed to individual turbulent scales and their mutual interaction.

Figure 6(a) first confirms that the decomposition cleanly separates Reynolds-number-invariant and Reynolds-number-dependent wall-pressure signatures. The inset shows that the inner-scale variance, $\overline{p^2_{w,i}}^+$, remains approximately invariant with Reynolds number, whereas the large-scale variance, $\overline{p^2_{w,L}}^+$, exhibits a logarithmic growth consistent with figures 3(a,b). Consequently, any Reynolds-number dependence of $\mathcal{S}_{p_w}$ must originate from the large scales and/or their interaction with the Reynolds-number-invariant inner scales. This hypothesis is examined explicitly in figure 6(b), which presents the decomposition of $\mathcal{S}_{p_w}$ for the HRNBLWT datasets. Consistent with analogous decompositions of near-wall turbulence (Mathis *et al.* 2011; Lozier *et al.* 2024), the contribution from $\overline{\overline{p_{w,L}^2 p_{w,i}}}$ remains negligibly small over the entire Reynolds-number range, while $\overline{\overline{p_{w,i}^3}}$ is essentially Reynolds-number invariant. The Reynolds-number dependence of $\mathcal{S}_{p_w}$ therefore arises almost entirely from two terms: the intrinsic skewness of the large-scale pressure fluctuations, $\overline{\overline{p_{w,L}^3}}$, and the nonlinear interaction term, $\overline{\overline{p_{w,i}^2 p_{w,L}}}$, both of which become progressively more negative as Reynolds number increases. Note here that although the absolute magnitudes of the decomposed terms depend on the choice of cutoff $T_c^+$, these Reynolds number trends remain robust over a wide range of $T_c^+$ (see figure 10 in Appendix B for confirmation).

Interestingly, the inner and large scales contribute to the total skewness with opposite signs, with the former positively skewed and the latter negatively skewed. Thus, the increasingly negative $\mathcal{S}_{p_w}$ does not arise because the inner scales themselves become more asymmetric. Instead, it reflects the growing influence of energetic large-scale pressure fluctuations and their nonlinear modulation of the statistically invariant inner scales. Together with the Reynolds-number-invariant and Reynolds-number-dependent behaviour of $\overline{p_{w,i}^2}^+$ and $\overline{p_{w,L}^2}^+$, respectively, these observations explain the transition from positive to negative $\mathcal{S}_{p_w}$ near $Re_\tau = 2414$ predicted by equation (3.1). An additional outcome of the decomposition in figure 6(b) is that it provides a physical explanation for the substantial scatter in $\mathcal{S}_{p_w}$ reported previously (figure 4a). Insufficient spatial resolution and/or inadequate Helmholtz resonance correction preferentially attenuate the positively skewed inner-scale contribution, whereas insufficient sampling duration primarily compromises convergence of the Reynolds-number-dependent large-scale and interaction terms. Consequently, all three experimental limitations can substantially alter the measured value of $\mathcal{S}_{p_w}$, even when the underlying turbulence is unchanged. This sensitivity is demonstrated using the HRNBLWT dataset at $Re_\tau \approx 14,700$ in figure 10 of Appendix B, where the time series could not be fully corrected for Helmholtz resonance.

Next, we examine the physical origin of the increasingly negative interaction term, $\overline{\overline{p_{w,i}^2 p_{w,L}}}$, which figure 6(b) identifies as one of the primary contributors to the Reynolds-number dependence of $\mathcal{S}_{p_w}$. A representative segment of the synchronously acquired signals at $Re_\tau \approx 11,300$ is shown in figure 6(c), including $p_{w,L}$, $p_{w,i}$ and the product $p_{w,i}^2 p_{w,L}$. Motivated by the classical picture of large-scale modulation of near-wall turbulence (Rao *et al.* 1971; Bandyopadhyay & Hussain 1984; Mathis *et al.* 2009), we investigate whether the Reynolds-number-dependent large-scale pressure fluctuations similarly modulate the Reynolds-number-invariant inner-scale wall-pressure fluctuations. The time traces reveal a clear asymmetry. Periods of negative large-scale wall pressure ($p_{w,L} < 0$) are preferentially associated with amplified inner-scale fluctuations, whereas periods with $p_{w,L} > 0$ correspond to comparatively weaker inner-scale activity. Consequently, $p_{w,i}^2 p_{w,L}$ assumes large negative values during negative large-scale pressure events, whereas the corresponding positive contribution during $p_{w,L} > 0$ remains substantially weaker. This asymmetry directly explains the negative sign of $\overline{\overline{p_{w,i}^2 p_{w,L}}}$. Moreover, because the large-scale wall-pressure fluctuations become both increasingly energetic and increasingly negatively skewed with increasing $Re_\tau$ (figures 6a,b), episodes of amplified inner-scale activity associated with $p_{w,L} < 0$ become progressively more frequent and intense. This provides a plausible physical explanation for the logarithmic Reynolds-number dependence of $\mathcal{S}_{p_w}$ described by equation (3.1). Independent support for this interpretation is provided by figure 6(d), which plots the ratio of the conditionally averaged inner-scale variance during intervals with $p_{w,L} < 0$ and $p_{w,L} > 0$, namely: $\langle p_{w,i}^2 \rangle_{p_{w,L}<0} / \langle p_{w,i}^2 \rangle_{p_{w,L}>0}$, where $\langle \cdot \rangle$ denotes conditional averaging. The ratio exceeds unity at all Reynolds numbers and increases systematically with $Re_\tau$, confirming that negative large-scale wall-pressure events become progressively more effective at amplifying the inner-scale fluctuations.

At first sight, this modulation mechanism appears opposite to the classical amplitude-modulation behaviour reported for near-wall velocity and wall-shear-stress fluctuations, where positive large-scale streamwise velocity fluctuations ($u_L$, obtained using the same cutoff $T_c^+ = 40$) preferentially enhance the inner-scale turbulence (Thomas & Bull 1983; Mathis *et al.* 2009, 2013; Pan & Kwon 2018). Figure 6(c), however, demonstrates that the two observations are entirely consistent. The synchronously acquired signals reveal a strong anti-correlation between $u_L$ measured at $z^+ \approx 15$ and $p_{w,L}$, with positive velocity fluctuations generally coinciding with negative wall-pressure fluctuations. As recognised previously (Thomas & Bull 1983; Lozier *et al.* 2025), positive $u_L$ fluctuations correspond to high-momentum sweep-like motions that locally accelerate the near-wall flow and are therefore associated with reduced wall pressure ($p_{w,L} < 0$). Consequently, the negative large-scale wall-pressure fluctuations identified here represent the pressure signature of the same energetic large-scale motions that are known to modulate the near-wall velocity field. The present results therefore extend the classical amplitude-modulation paradigm from velocity to wall-pressure fluctuations, consistent with Tsuji *et al.* (2016). This interpretation supports the hypothesis that the Reynolds-number dependence of $\mathcal{S}_{p_w}$ is correlated with the progressive energisation and broadening of the hierarchy of large-scale motions with increasing Reynolds number (Marusic *et al.* 2015; Deshpande *et al.* 2025). Having established this physical picture, we next quantify the stochastic coupling between $u$ and $p_w$ to identify the wall-normal regions and turbulent scales associated with the observed behaviour.

### 3.3. *Linear and quadratic coherence between streamwise velocity and wall-pressure*

The analysis in §3.2 established that the Reynolds-number variation of $\mathcal{S}_{p_w}$ arises from the increasing influence of large-scale wall-pressure fluctuations and their nonlinear interaction with statistically invariant inner-scale fluctuations. This behaviour is reminiscent of the well-established amplitude-modulation mechanism in wall turbulence, whereby energetic large-scale motions in the logarithmic region linearly superimpose onto, and nonlinearly modulate, the near-wall small-scale turbulence (Rao *et al.* 1971; Bandyopadhyay & Hussain 1984; Mathis *et al.* 2009, 2011; Lozier *et al.* 2024). Similar interactions have also been reported for wall quantities, including near-wall pressure (Thomas & Bull 1983; Tsuji *et al.* 2016) and wall-shear stress fluctuations (Mathis *et al.* 2013; Pan & Kwon 2018), although the interpretation for the former is not as straightforward given it is an integral across multiple pressure sources in the semi-infinite space above the wall. The key question addressed here is therefore limited to: which streamwise velocity flow scales ($T^+$) and wall-normal regions ($z^+$) are *associated* with the Reynolds-number evolution of the wall-pressure statistics?

To answer this, we examine the linear ($\gamma_{up_w}^2$) and quadratic ($\gamma_{up_w^2}^2$) coherence spectra between $u$ and $p_w$, which quantify the scale-resolved stochastic relationship between the overlying velocity field and wall-pressure fluctuations across all $z^+$ and $T^+$. Notably, $\gamma_{up_w}^2$ quantifies the scale-by-scale linear correlation between $u(z)$ and $p_w$, whereas $\gamma_{up_w^2}^2$ quantifies the extent to which large-scale velocity fluctuations at $z$ are coupled nonlinearly with the wall-pressure fluctuations through an amplitude-modulation-type mechanism (Baars *et al.*

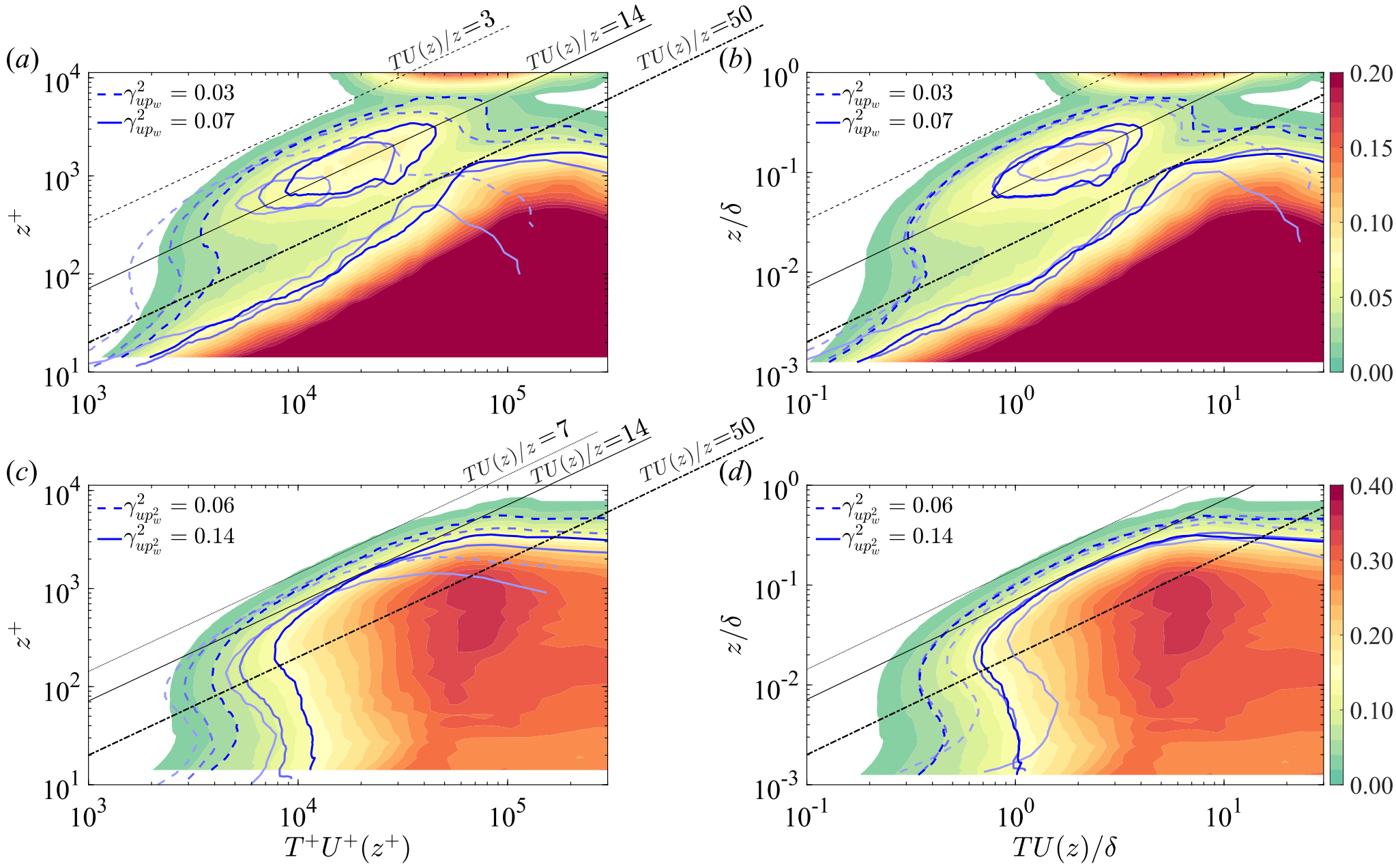

Figure 7: (a-d) Linear ($\gamma^2_{up_w}$) and quadratic ($\gamma^2_{up^2_w}$) coherence spectrogram between (a,b) $u$ and $p_w$, and (c,d) $u$ and $p^2_w$, plotted as a function of (a,c) inner- and (b,d) outer-scaled time scales and distance from the wall. The coloured contours are for $Re_\tau \approx 11,300$, while the blue iso-contours are for $Re_\tau \approx 5000$, 8000 and 11,300, with an increase in colour intensity corresponding to an increase in $Re_\tau$ (table 1). Black dashed, dotted, solid and dash-dotted lines denote self-similar scalings associated with the data: $TU(z) = 3z$, $TU(z) = 7z$, $TU(z) = 14z$ and $TU(z) = 50z$, respectively.

2024; Dacome *et al.* 2025). These quantities are respectively defined as:

$$\gamma^2_{up_w}(z^+;T^+) = \frac{\left|\{\widetilde{U}(z^+;T^+)\,\widetilde{P}^*_w(T^+)\}\right|^2}{\{\widetilde{U}(z^+;T^+)\,\widetilde{U}^*(z^+;T^+)\}\{\widetilde{P}_w(T^+)\,\widetilde{P}^*_w(T^+)\}} \quad \text{and} \tag{3.3}$$

$$\gamma^2_{up^2_w}(z^+;T^+) = \frac{\left|\{\widetilde{U}(z^+;T^+)\,\widetilde{P^2_w}^*(T^+)\}\right|^2}{\{\widetilde{U}(z^+;T^+)\,\widetilde{U}^*(z^+;T^+)\}\{\widetilde{P^2_w}(T^+)\,\widetilde{P^2_w}^*(T^+)\}}, \tag{3.4}$$

where $\{\cdot\}$ denotes ensemble averaging, $*$ denotes complex conjugation, and $\widetilde{U}$, $\widetilde{P}_w$ and $\widetilde{P^2}_w$ denote the Fourier transforms of $u$, $p_w$ and $p^2_w$, respectively. Here, $p^2_w = [p^2_w]_t - \overline{[p^2_w]_t}$, where $[p^2_w]_t$ represents the time series of squared wall-pressure, prior to subtraction of its mean (Dacome *et al.* 2025). Using simultaneously acquired $u$ and $p_w$ signals from the HRNBLWT, together with comparable datasets from LES and SLTEST, we identify the turbulent flow motions responsible for the Reynolds-number variation of $\gamma^2_{up_w}$ and $\gamma^2_{up^2_w}$ in figures 7 and 8, thereby respectively linking the evolution of $\overline{p^2_w}^+$ and $\mathcal{S}_{p_w}$ to specific wall-normal regions and turbulent flow scales.

Figures 7(a,b) and 7(c,d) present the two-dimensional distributions of $\gamma^2_{up_w}$ and $\gamma^2_{up^2_w}$, respectively, obtained from the HRNBLWT datasets across $5000 \lesssim Re_\tau \lesssim 11,300$. Figures 8(a,b) and 8(c,d) complement these results by comparing selected wall-normal locations across LES, HRNBLWT and SLTEST datasets spanning $10^3 \lesssim Re_\tau \lesssim 10^6$.

Throughout figures 7 and 8, the abscissa is expressed as $TU(z)$, following Baars *et al.* (2024) and Dacome *et al.* (2025), thereby facilitating comparison across datasets and with the proposed self-similar scaling laws. Taylor's hypothesis is well justified over the intermediate and large scales considered here (see figure 14 of Deshpande *et al.* 2023), while recent measurements have further confirmed that large-scale wall-pressure fluctuations convect at velocities comparable to their associated turbulent motions (Deshpande *et al.* 2026; Butt *et al.* 2026). Overall, the coherence spectra recover the principal scaling characteristics reported previously (Baars *et al.* 2024; Dacome *et al.* 2025), while revealing new Reynolds-number-dependent behaviour that is directly associated with the evolution of $\overline{p_w^2}^+$ and $\mathcal{S}_{p_w}$.

We first consider the linear coherence, $\gamma^2_{up_w}$. Consistent with previous studies (Baars *et al.* 2024; Dacome *et al.* 2025), negligible coherence exists over the small-scale regime ($TU/z \lesssim 3$), indicated by the dashed black lines in figures 7(a,b). Beyond this threshold, however, two dynamically distinct coherence regimes emerge, as highlighted by the $\gamma^2_{up_w} = 0.07$ isocontours. The first regime spans approximately $3 \lesssim TU/z \lesssim 50$ and occupies both the logarithmic and outer regions ($z^+ \gtrsim 100$). Within this intermediate-scale regime, the coherence contours closely follow the Reynolds-number-invariant scaling $TU \approx 14z$, consistent with the self-similar attached-eddy hierarchy (Baars *et al.* 2024; Dacome *et al.* 2025; Deshpande *et al.* 2025). More specifically, these motions correspond to Townsend's (1976) 'active' motions, namely those primarily responsible for the local Reynolds shear stress (Deshpande *et al.* 2025). Here, although the local coherence remains approximately Reynolds-number invariant, the self-similar hierarchy spans progressively larger ranges of wall-normal positions and timescales as $Re_\tau$ increases. Consequently, the cumulative linear correlation between the attached-eddy hierarchy and the wall-pressure field increases despite the approximately invariant local coherence. A second coherence regime appears for $TU/z \gtrsim 50$ and is concentrated primarily below $z \lesssim 0.15\delta$, where it is associated with relatively large attached eddies (relative to the local wall-normal position) and turbulent superstructures. This regime corresponds with Townsend's (1976) 'inactive' motions, which contribute only weakly to the local Reynolds shear stress (Deshpande *et al.* 2025). Unlike the intermediate-scale regime, however, $\gamma^2_{up_w}$ in this regime increases systematically with Reynolds number.

The distinct Reynolds-number behaviour of these two coherence regimes becomes even clearer in figures 8(a,b), which compare $\gamma^2_{up_w}$ at fixed wall-normal locations ($z^+ \approx 350$ and 1500) across $10^3 \lesssim Re_\tau \lesssim 10^6$. The intermediate-scale regime remains centred around the attached-eddy scaling of $TU \approx 14z$ and exhibits little variation over almost three decades of Reynolds number, consistent with earlier observations (Baars *et al.* 2024; Dacome *et al.* 2025). By contrast, the large-scale regime exhibits pronounced $Re_\tau$-growth for $\gamma^2_{up_w}$ compared at the same $z^+$, indicating enhanced correlation between the large-scale streamwise velocity and wall-pressure field for increasing $Re_\tau$. To confirm this interpretation, figures 8(a,b) also compare $\gamma^2_{up_w}$ with the coherence spectrum $\gamma^2_{u_L p_{w,L}}$ obtained by respectively replacing $u$ and $p_w$ with $u_L$ and $p_{w,L}$ in equation (3.3). Since $u_L = u$ ($T^+ > 40$) and $p_{w,L} = p_w$ ($T^+ > 40$), the close agreement between the two spectra at large scales is expected. Nevertheless, their near-perfect collapse demonstrates that the Reynolds-number growth of $\overline{p_w^2}^+$ is associated with the increasingly strong correlation between the self-similar attached-eddy hierarchy and the $\delta$-scaled turbulent superstructures with the wall-pressure field.

Next, we examine whether these same motions are also associated with the evolution of wall-pressure skewness through nonlinear amplitude modulation (*i.e.*, the quadratic coherence). Consistent with previous observations, negligible quadratic coherence exists for $TU/z \lesssim 7$, indicated by the dotted black lines in figures 7(c,d). Relative to $\gamma^2_{up_w}$, the coherence is shifted towards larger scales, reflecting the fact that squaring $p_w$ introduces lower-frequency content associated with temporal variations in the intensity of the high-frequency wall-pressure fluctuations (Baars *et al.* 2024; Dacome *et al.* 2025). Remarkably, $\gamma^2_{up^2_w}$ exhibits the same two coherence regimes identified previously for $\gamma^2_{up_w}$. This demonstrates that the streamwise velocity motions, which are linearly correlated with the wall-pressure field, are also those associated with its nonlinear amplitude modulation. This similarity is evident in both the two-dimensional coherence maps (figures 7c,d) and the fixed-$z^+$ comparisons (figures 8c,d). Within the intermediate-scale regime ($7 \lesssim TU/z \lesssim 50$), the coherence remains closely centred about the self-similar scaling $TU \approx 14z$ over the entire Reynolds-number range considered. This indicates that the role played by the attached-eddy hierarchy, towards the amplitude modulation of wall-pressure fluctuations, is approximately Reynolds-number invariant. Nevertheless, because the attached-eddy hierarchy broadens over progressively larger wall-normal distances and scales with increasing $Re_\tau$, its cumulative contribution to the amplitude modulation also increases. Similarly, in the large-scale regime ($TU/z \gtrsim 50$), $\gamma^2_{up^2_w}$ increases systematically with Reynolds number, indicating progressively stronger nonlinear coupling between the largest velocity motions and the fluctuating intensity of the wall-pressure field. The strongest values of $\gamma^2_{up^2_w}$ originate predominantly within the logarithmic region ($2.6\sqrt{Re_\tau} \lesssim z^+ \lesssim 0.15Re_\tau$; Wei *et al.* 2005), demonstrating that the attached-eddy hierarchy and turbulent superstructures are associated most significantly with the nonlinear modulation of wall-pressure fluctuations. This observation establishes the physical link between the Reynolds-number evolution of $\overline{p^2_w}^+$ (§3.1) and $\mathcal{S}_{p_w}$ (§3.2).

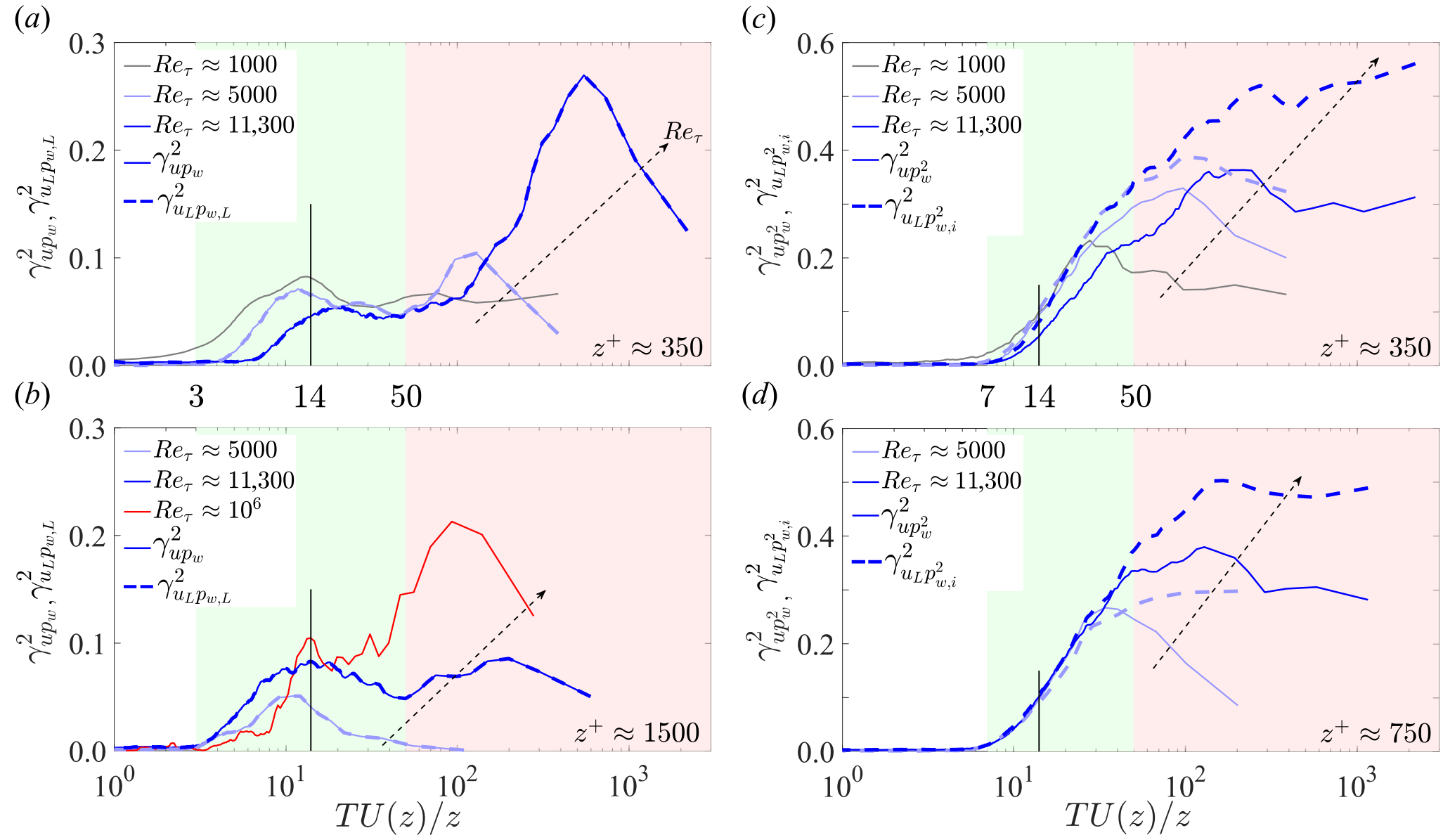

Figure 8: (a,b) Linear and (c,d) quadratic coherence spectra plotted against a wall-scaled abscissa at $z^+ \approx$ (a,c) 350, (b) 1500, (d) 750 across $10^3 \lesssim Re_\tau \lesssim 10^6$. While solid blue lines represent coherence between (a,c) $u$ and $p_w$, and (c,d) $u$ and $p_w^2$, dashed blue lines represent the same between (a,b) $u_L$ and $p_{w,L}$, and (c,d) $u_L$ and $p^2_{w,i}$. Solid black line denotes $TU(z) = 14z$, while the green and red shading represents flow scales associated with intermediate ($3z \lesssim TU \lesssim 50z$) and large-scale ($TU > 50z$) motions, respectively.

To connect these observations directly with the classical amplitude-modulation framework (Mathis *et al.* 2009, 2011), we finally isolate the modulation pathway by directly evaluating the nonlinear interaction between large-scale velocity fluctuations (that are linearly correlated with the large-scale wall-pressure, $p_{w,L}$) and the inner-scale wall-pressure fluctuations ($p_{w,i}$). Specifically, we compute the quadratic coherence spectrum ($\gamma^2_{u_L p^2_{w,i}}$) for the HRNBLWT datasets using an equation analogous to (3.4), replacing $u \rightarrow u_L$ and $p^2_w \rightarrow p^2_{w,i}$. The resulting spectra are compared with $\gamma^2_{u p^2_w}$ in figures 8(c,d). Notably, $\gamma^2_{u_L p^2_{w,i}}$ reproduces the same intermediate-scale scaling ($TU \approx 14z$), the same Reynolds-number growth within the large-scale regime, and an overall spectral shape closely matching that of $\gamma^2_{u p^2_w}$. The remaining differences in magnitude are expected because figure 6(b) demonstrated that $\overline{p^2_{w,i} p_{w,L}}$ represents only one of the two $Re_\tau$-dependent contributions to the total wall-pressure skewness. Nevertheless, the close agreement between the two coherence spectra confirms that the Reynolds-number variation of wall-pressure skewness is associated with the same logarithmic-region motions correlated with the the Reynolds-number growth of the wall-pressure variance.

## 4. Concluding remarks

This study presents an investigation of wall-pressure fluctuations beneath zero-pressure-gradient turbulent boundary layers across $O(10^3) \lesssim Re_\tau \lesssim O(10^4)$ using new, well-resolved laboratory measurements in the Melbourne boundary layer wind tunnel. They are complemented by atmospheric surface-layer measurements approaching $Re_\tau \sim O(10^6)$ and published large-eddy simulations at $Re_\tau \sim O(10^3)$. By combining broadband spectral analysis, second- and third-order statistics, and scale-resolved coherence analysis between $u$ and $p_w$, the present work addresses three central questions: the Reynolds-number dependence of the wall-pressure spectrum, the evolution of wall-pressure skewness, and the turbulent scales and wall-normal regions most strongly associated with their $Re_\tau$-trends.

The spectral analysis for ZPG TBL showed that, unlike turbulent channel and pipe flows (Panton *et al.* 2017; Yu *et al.* 2022; Baars *et al.* 2024; Pirozzoli & Wei 2025), the small-scale region of the inner-scaled wall-pressure spectrum exhibits negligible Reynolds-number growth once measurement limitations are carefully understood and removed. This result supports the view that the small-scale Reynolds-number growth reported in internal flows is not universal across canonical wall-bounded turbulence, but is instead likely influenced by geometric confinement and/or pressure contributions from turbulent motions located beyond the channel or pipe centreline. This is an important caveat when modelling inner-scaled $p_w$ statistics for internal flows based on external-flow measurements, or vice versa. At intermediate and large scales, however, the ZPG TBL exhibits Reynolds-number growth consistent with previous observations in canonical wall flows, reinforcing the association of inertia-dominated motions, including the attached-eddy hierarchy and turbulent superstructures, with the energetic wall-pressure fluctuations.

A key contribution of the present study is the establishment of a clear Reynolds-number dependence of wall-pressure skewness, $\mathcal{S}_{p_w}$, for the case of a ZPG TBL. While the wall-pressure probability density function remains approximately symmetric at low-to-moderate Reynolds numbers, increasing $Re_\tau$ produces progressively stronger negative skewness through an increasing frequency of moderately strong negative-pressure events, rather than increasingly rare extreme events. The scale-based decomposition of $\mathcal{S}_{p_w}$ further revealed that this behaviour arises primarily from two Reynolds-number-dependent mechanisms: (a) the increasingly negative skewness of the large-scale wall-pressure component, and (b) its nonlinear interaction with statistically invariant inner-scale wall-pressure fluctuations. In contrast, the intrinsic skewness of the inner-scale wall-pressure fluctuations remains approximately Reynolds-number invariant and is, in fact, predominantly positive. These observations also provide a plausible physical explanation for the empirically observed logarithmic variation of $\mathcal{S}_{p_w}$ with $Re_\tau$, whereby increasingly energetic and negatively skewed large-scale wall-pressure fluctuations, together with their strengthening nonlinear interaction with the statistically invariant inner scales, progressively dominate the total skewness. These observations provide direct evidence that higher-order wall-pressure statistics become increasingly important at high Reynolds numbers, with implications for stochastic-estimation frameworks and predictive models that otherwise assume near-Gaussian wall-pressure statistics, as in previous low-$Re_\tau$ studies (Naguib *et al.* 2001; Baars *et al.* 2024).

The coherence analysis identified the turbulent motions associated with these Reynolds-number-dependent observations. Both the linear ($\gamma^2_{up_w}$) and quadratic ($\gamma^2_{up^2_w}$) coherence spectra exhibited two dynamically distinct coherence regimes separated approximately by $TU/z \approx 50$. Within the intermediate-scale regime ($7 \lesssim TU/z \lesssim 50$ for quadratic coherence, with linear coherence extending down to $TU/z \approx 3$), the local coherence collapses under distance-from-the-wall scaling and remains approximately Reynolds-number invariant, consistent with coupling between wall pressure and the geometrically self-similar attached-eddy hierarchy, populating the logarithmic region. Nevertheless, because this hierarchy broadens over progressively larger wall-normal extents and turbulent scales in inner-scaled coordinates, its cumulative contribution to both linear correlation and nonlinear amplitude modulation increases with Reynolds number. In contrast, the larger-scale regime ($TU/z \gtrsim 50$), associated with relatively large attached eddies and turbulent superstructures, exhibits explicit Reynolds-number growth in both linear and quadratic coherence, demonstrating progressively stronger coupling between these motions and the wall-pressure field. Together, these observations reveal that the Reynolds-number growth of wall-pressure variance and skewness is associated with the same logarithmic-region motions (*i.e.*, the attached-eddy hierarchy and turbulent superstructures).

Finally, the aforementioned physical insights were reached only after careful consideration of several experimental aspects found to be critical for obtaining accurate wall-pressure statistics. First, statistical convergence of higher-order moments requires substantially longer acquisition times than those needed for converged spectra. While second-order statistics converge for acquisition durations of $TU_\infty/\delta \gtrsim O(10^4)$, reliable estimation of skewness requires records approaching $TU_\infty/\delta \sim O(10^5)$ or larger. This disparity explains why Reynolds-number trends in $\mathcal{S}_{p_w}$ have remained ambiguous in many previous studies and underscores the importance of explicitly assessing statistical convergence when interpreting higher-order wall-pressure statistics. Second, the interaction between sensor geometry and Helmholtz resonance requires careful treatment. The present work demonstrates that merely correcting the resonance peak is insufficient when the resonance bandwidth overlaps the energetic portion of the wall-pressure spectrum (see Appendices A and B). Reliable high-Reynolds-number measurements therefore benefit from experimental designs that place the resonance frequency well beyond the energetic spectral range before any correction is applied. More broadly, these findings emphasise that instrumentation design should be regarded as integral to wall-pressure experiments, with post-processing corrections serving as a complement rather than a substitute.

**Acknowledgements**

R. Deshpande is grateful to the University of Melbourne and RMIT University for financial support from the Melbourne Postdoctoral Fellowship and the Vice Chancellor's Fellowship, respectively. R. Deshpande and I. Marusic also gratefully acknowledge funding from the Office of Naval Research (ONR) and ONR Global; Grant No. N62909-23-1-2068. I. Marusic is supported by funding from a Defense Advanced Research Projects Agency (DARPA) grant.

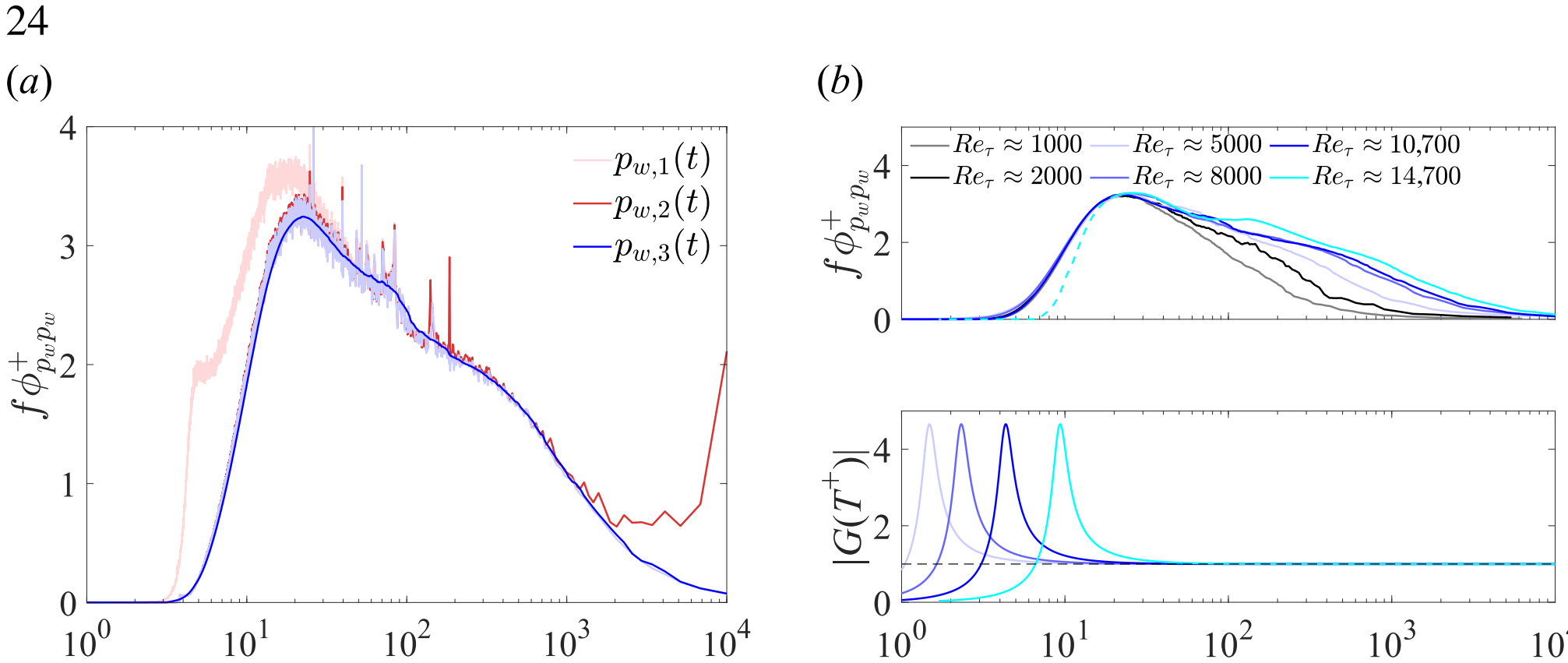

Figure 9: (a) Premultiplied frequency spectra of $p_w$ at $Re_\tau \approx 10,700$ obtained from raw wall-pressure signal, $p_{w,1}$, after Helmholtz resonance corrections, $p_{w,2}$, and after facility noise treatment using Wiener noise-cancelling filter, $p_{w,3}$. (b) Final corrected premultiplied frequency spectra of $p_w$ at various $Re_\tau$ (top) and their corresponding transfer functions (bottom). Data across $5000 \lesssim Re_\tau \lesssim 14{,}700$ are from the present study, while $Re_\tau \approx 1000$ and 2000 are from the LES dataset (Eitel-Amor *et al.* 2014).

## Conflict of interest

The authors declare that they have no conflict of interest.

## Data repository

Wall-pressure statistics from the current study will be made publicly available online after publication of the manuscript. Researchers interested in gaining access to statistics before publication can contact the corresponding author.

## Appendix A. Wall-pressure spectra corrected for Helmholtz resonance and facility noise

Figure 9(a) compares the premultiplied wall-pressure spectra at $Re_\tau \approx 10,700$ before and after application of the Helmholtz-resonance and facility-noise corrections. Following the two corrections, a bandwidth-moving filter was applied to the resulting spectra, $f\phi^+_{p_wp_w}$, using a window of $T^+ \simeq \pm 20\%$. The corrected spectra for all Reynolds numbers considered in the present study are shown in the top portion of figure 9(b), while the corresponding transfer-function magnitude expressed as a function of $T^+$ in the bottom half. Also included is the spectrum estimated based on the measurement at $Re_\tau \approx 14,700$, to demonstrate the limitations associated with Helmholtz-resonance corrections at higher $Re_\tau$. In the present experiments, higher Reynolds numbers were achieved by increasing the freestream velocity, which resulted in progressively smaller viscous time scales $(\nu/U_\tau^2)$. Consequently, the inner-normalised transfer function moves towards the larger time scales (*i.e.*, lower frequencies) in the spectra, as depicted in figure 9b. For the $Re_\tau \approx 14,700$ case, the spectrum is influenced for approximately $T^+ \lesssim 25$, making this portion of the spectrum unreliable (as indicated by dashed lines in figure 9b), even after applying the Helmholtz correction. Although the resonance correction is theoretically well-founded, previous studies (Dacome *et al.* 2025) have shown that the transfer function may differ slightly when the resonator is excited by wall-bounded turbulence rather than acoustic waves alone. This difference is owing to changes in the effective end correction and damping characteristics. In contrast, for the $Re_\tau \approx 5000$, 8000, and 10,700 datasets, the resonance bandwidth remains largely outside the energetic portion of the spectrum, enabling reliable recovery of the complete spectral range. Nevertheless, application of the correction remains necessary to account for residual resonance effects and ensure consistency across all datasets.

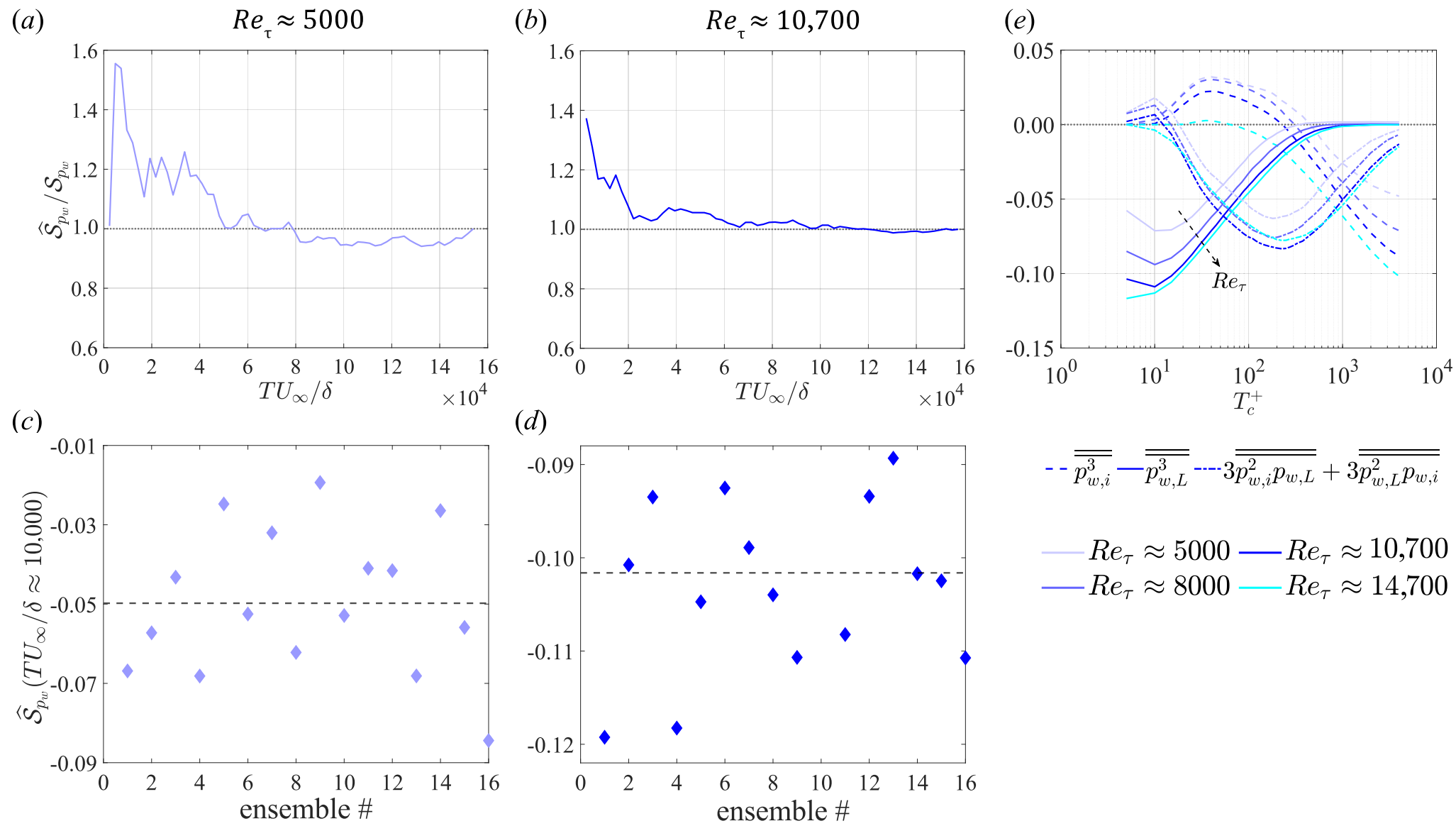

Figure 10: (a,b) Statistical convergence of the skewness of wall-pressure for increasing time series length, $TU_\infty/\delta$. Here, $\widehat{S}_{p_w}$ is the estimated for $0 < TU_\infty/\delta < 160{,}000$ while $S_{p_w}$ is the mean skewness of $p_w$ obtained on considering the full time series length, $TU_\infty/\delta \approx 160{,}000$. (c-d) Skewness of wall-pressure for sixteen different ensembles, obtained on dividing the full $p_w$-time series into shorter segments of $TU_\infty/\delta \approx 10000$ each. The dashed line represents $S_{p_w}$ based on $TU_\infty/\delta \approx 160{,}000$. (e) Scale-based decomposition of $S_{p_w}$ into its 4 different subcomponents following (3.2), for varying viscous-scaled cutoff time scale ($T_c^+$) and $Re_\tau$.

## Appendix B. Statistical convergence of wall-pressure skewness and its scale-based decomposition

The statistical convergence of wall-pressure skewness was assessed by evaluating over progressively increasing eddy-turnover times, $0 < TU_\infty/\delta \lesssim 160{,}000$, denoted as $\widehat{S}_{p_w}$, and comparing it to the final converged estimate, $S_{p_w}$, obtained using the full record length ($TU_\infty/\delta \approx 160{,}000$). Figures 10(a,b) show $\widehat{S}_{p_w}/S_{p_w}$ as a function of $TU_\infty/\delta$ for $Re_\tau \approx 5000$ and 10,700, respectively. The ratio exhibits significant fluctuations at short eddy-turnover times, but converges towards unity only when $TU_\infty/\delta \gtrsim 10^5$, confirming that reliable estimation of the third-order moment of $p_w$ requires substantially long acquisition times. To illustrate the scatter associated with unconverged estimates, the full $p_w$ time series was divided into 16 ensembles, each corresponding to $TU_\infty/\delta \approx 10{,}000$. The skewness values obtained from these individual ensembles, denoted as $\widehat{S}_{p_w}(TU_\infty/\delta \approx 10{,}000)$, are shown in figures 10(c,d) for $Re_\tau \approx 5000$ and 10,700, respectively. Substantial scatter is observed relative to the converged mean skewness, $S_{p_w}$, with deviations reaching more than ±20%. This sensitivity to sampling duration provides a likely explanation for much of the scatter (*i.e.*, absence of any $Re_\tau$-trend) in the $S_{p_w}$ values reported in the literature (figure 4a).

Figure 10(e) quantifies the sensitivity of the scale-based decomposition of $S_{p_w}$ (equa-tion 3.2) to the cutoff timescale $T_c^+$. For the well-resolved datasets spanning $5000 \lesssim Re_\tau \lesssim 10,700$, the Reynolds-number trends of $\overline{\overline{p_{w,L}^3}}$ and $\overline{\overline{p_{w,L}^2 p_{w,i} + p_{w,i}^2 p_{w,L}}}$ are largely insensitive to the choice of $T_c^+$. In contrast, the small-scale skewness, $\overline{\overline{p_{w,i}^3}}$, is quasi-invariant only for $T_c^+ \lesssim 40$ and becomes Reynolds-number dependent as $T_c^+$ increases to include a broader scale range. The data at $Re_\tau \approx 14,700$ represent a special case excluded from the main analysis because the influence of Helmholtz resonance could not be fully corrected (figure 9). The under-resolved small-scale end of the $p_w$ spectrum strongly affects the skewness estimate, resulting in $\overline{\overline{p_{w,i}^3}} \approx 0$ for $T_c^+ \lesssim 40$, in clear contrast to the quasi-$Re_\tau$-invariant behaviour observed for the lower-$Re_\tau$ cases.

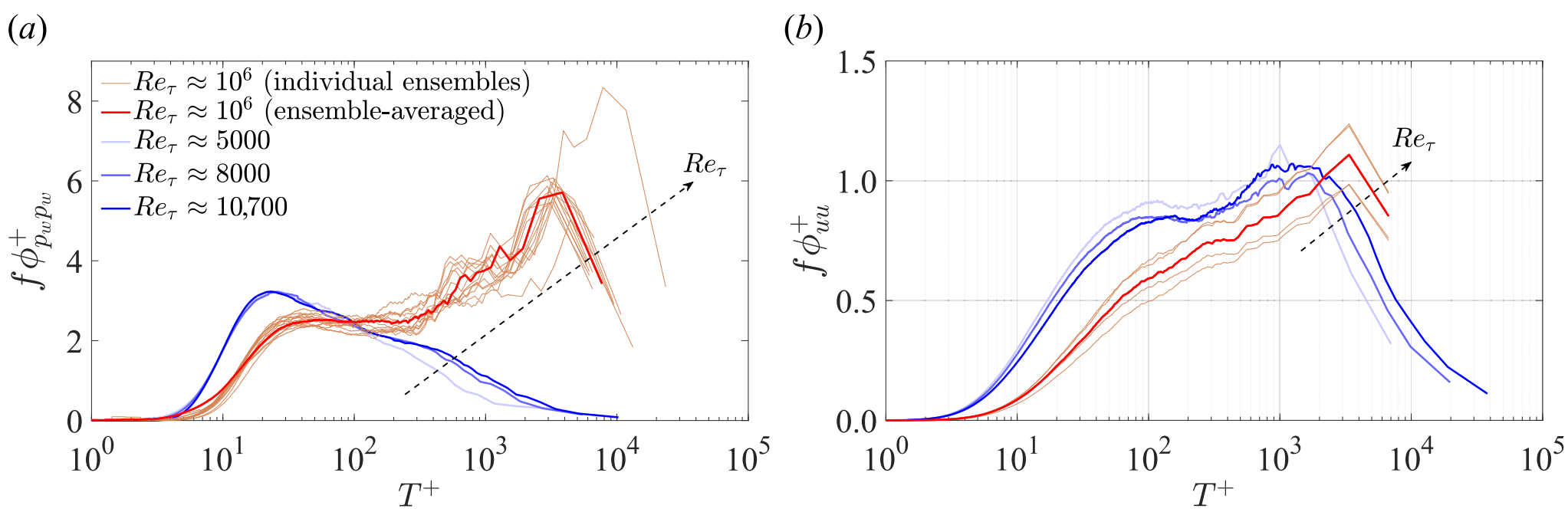

Figure 11: Premultiplied frequency spectra of (a) $p_w$ and (b) $u$ at $Re_\tau \approx 10^6$, obtained from various SLTEST data acquired during the 2003-2004 campaigns using different microphones and hotwire sensors (orange lines). The ensemble-averaged estimate is shown in red and compared against well-resolved spectra from HRNBLWT for $5000 \lesssim Re_\tau \lesssim 10,700$.

## Appendix C. Ensemble-averaged spectra obtained from independent SLTEST datasets

As discussed in §2.2, the SLTEST dataset comprises hot-wire and microphone measurements acquired on different days during the 2003–2004 campaigns, using different sets of hot-wire sensors and wall-mounted microphones. The premultiplied frequency spectra obtained from each individual ensemble are plotted using orange lines in figures 11(a,b) for the $p_w$ and $u$-spectra, respectively. The ensemble-averaged spectra are shown in red correspond to the spectra used in the main analysis in figures 3(b,d).

REFERENCES

Abe, H., Matsuo, Y. & Kawamura, H. 2005 A DNS study of Reynolds-number dependence on pressure fluctuations in a turbulent channel flow. In *Fourth International Symposium on Turbulence and Shear Flow Phenomena*. Begel House Inc.

Andreopoulos, J. & Agui, J. H. 1996 Wall-vorticity flux dynamics in a two-dimensional turbulent boundary layer. *J. Fluid Mech.* **309**, 45–84.

Baars, W. J., Dacome, G. & Lee, M. 2024 Reynolds-number scaling of wall-pressure–velocity correlations in wall-bounded turbulence. *J. Fluid Mech.* **981**, A15.

Bandyopadhyay, P. R. & Hussain, A. K. M. F. 1984 The coupling between scales in shear flows. *Phys. Fluid* **27** (9), 2221–2228.

Butt, H., Damani, S., Devenport, W. J. & Todd Lowe, K. 2026 Identification of sources of wall pressure fluctuations using space–time pressure–velocity correlations. *AIAA J.* pp. 1–14.

Corcos, G. M. 1964 The structure of the turbulent pressure field in boundary-layer flows. *J. Fluid Mech.* **18**, 353–378.

Dacome, G., Lazzarini, L., Talamelli, A., Bellani, G. & Baars, W. J. 2025 Scaling of wall-pressure–velocity correlations in high-Reynolds-number turbulent pipe flow. *J. Fluid Mech.* **1013**, A48.

Deshpande, R., Hassanein, A. & Baars, W. J. 2026 Convection velocities and velocity coupling of outer-scaled wall-pressure fluctuations in canonical turbulent boundary layers. *Phys. Rev. Fluids* **11**, 064612.

Deshpande, R., de Silva, C. M. & Marusic, I. 2023 Evidence that superstructures comprise self-similar coherent motions in high Reynolds number boundary layers. *J. Fluid Mech.* **969**, A10.

Deshpande, R., Vinuesa, R., Klewicki, J. & Marusic, I. 2025 Active and inactive contributions to the wall pressure and wall-shear stress in turbulent boundary layers. *J. Fluid Mech.* **1003**, A24.

Eitel-Amor, G., Örlü, R. & Schlatter, P. 2014 Simulation and validation of a spatially evolving turbulent boundary layer up to $Re_\theta = 8300$. *Int. J. Heat Fluid Flow* **47**, 57–69.

Farabee, T. M. & Casarella, M. J. 1991 Spectral features of wall pressure fluctuations beneath turbulent boundary layers. *Phys. Fluids* **3**, 2410–2420.

Fritsch, D. J., Vishwanathan, V., Todd Lowe, K. & Devenport, W. J. 2022 Fluctuating pressure beneath smooth wall boundary layers in nonequilibrium pressure gradients. *AIAA J.* **60** (8), 4725–4743.

Ghaemi, S. & Scarano, F. 2013 Turbulent structure of high-amplitude pressure peaks within the turbulent boundary layer. *J. Fluid Mech.* **735**, 381–426.

Gibeau, B. & Ghaemi, S. 2021 Low- and mid-frequency wall-pressure sources in a turbulent boundary layer. *J. Fluid Mech.* **918**, A18.

Gravante, S. P., Naguib, A. M., Wark, C. E. & Nagib, H. M. 1998 Characterization of the pressure fluctuations under a fully developed turbulent boundary layer. *AIAA J.* **36**, 1808–1816.

Hayes, M. H. 1996 *Statistical digital signal processing and modeling*. John Wiley & Sons.

Hutchins, N., Chauhan, K., Marusic, I., Monty, J. & Klewicki, J. 2012 Towards reconciling the large-scale structure of turbulent boundary layers in the atmosphere and laboratory. *Boundary-Layer Meteorol.* **145**, 273–306.

Jiang, C., de Silva, C., Doolan, C. & Moreau, D. 2025 Design and characterisation of an open-jet pressure gradient test rig for an aeroacoustic wind tunnel. *Appl. Acoust.* **227**, 110214.

Jiménez, J. 2004 Turbulent flows over rough walls. *Annu. Rev. Fluid Mech.* **36** (1), 173–196.

Klewicki, J. C., Priyadarshana, P. J. A. & Metzger, M. M. 2008 Statistical structure of the fluctuating wall pressure and its in-plane gradients at high Reynolds number. *J. Fluid Mech.* **609**, 195–220.

Knoop, M. W., Hassanein, A. & Baars, W. J. 2026 Development and characterisation of a turbulent boundary layer facility at the Delft University of Technology. *Aerosp. Sci. Technol.* **168**, 110972.

Lamballais, E., Lesieur, M. & Métais, O. 1997 Probability distribution functions and coherent structures in a turbulent channel. *Phys. Rev. E* **56** (6), 6761.

Lee, J. H. & Sung, H. J. 2013 Comparison of very-large-scale motions of turbulent pipe and boundary layer simulations. *Phys. Fluids* **25** (4).

Lozier, M., Deshpande, R., Zarei, A., Lindić, L., Rowin, W. A. & Marusic, I. 2025 Defining the mean turbulent boundary layer thickness based on streamwise velocity skewness. *J. Fluid Mech.* **1021**, A19.

Lozier, M., Marusic, I. & Deshpande, R. 2024 Revisiting amplitude modulation in non-canonical wall-turbulence through high-Reynolds number experimental data. *Phys. Rev. Fluids* **9** (12), 124602.

Luhar, M, Sharma, AS & McKeon, BJ 2014 On the structure and origin of pressure fluctuations in wall turbulence: predictions based on the resolvent analysis. *J. Fluid Mech.* **751**, 38–70.

Marusic, I., Chauhan, K. A., Kulandaivelu, V. & Hutchins, N. 2015 Evolution of zero-pressure-gradient boundary layers from different tripping conditions. *J. Fluid Mech.* **783**, 379–411.

Mathis, R., Hutchins, N. & Marusic, I. 2009 Large-scale amplitude modulation of the small-scale structures in turbulent boundary layers. *J. Fluid Mech.* **628**, 311–337.

Mathis, R., Marusic, I., Chernyshenko, S. I. & Hutchins, N. 2013 Estimating wall-shear-stress fluctuations given an outer region input. *J. Fluid Mech.* **715**, 163–180.

Mathis, R., Marusic, I., Hutchins, N. & Sreenivasan, K. R. 2011 The relationship between the velocity skewness and the amplitude modulation of the small scale by the large scale in turbulent boundary layers. *Phys. Fluid* **23** (12).

Metzger, M., McKeon, B. J. & Holmes, H. 2007 The near-neutral atmospheric surface layer: turbulence and non-stationarity. *Philos. Trans. R. Soc. Lond. A* **365**, 859–876.

Monty, J. P., Hutchins, N., Ng, H. C. H., Marusic, I. & Chong, M. S. 2009 A comparison of turbulent pipe, channel and boundary layer flows. *J. Fluid Mech.* **632**, 431–442.

Naguib, A. M., Wark, C. E. & Juckenhöfel, O. 2001 Stochastic estimation and flow sources associated with surface pressure events in a turbulent boundary layer. *Phys. Fluids* **13**, 2611–2626.

Naka, Y., Stanislas, M., Foucaut, J. M., Coudert, S., Laval, J. P. & Obi, S. 2015 Space–time pressure–velocity correlations in a turbulent boundary layer. *J. Fluid Mech.* **771**, 624–675.

Pan, C. & Kwon, Y. 2018 Extremely high wall-shear stress events in a turbulent boundary layer. In *J. Phys.: Conf. Ser.*, , vol. 1001, p. 012004. IOP Publishing.

Panton, R. L., Lee, M. & Moser, R. D. 2017 Correlation of pressure fluctuations in turbulent wall layers. *Phys. Fluids* **2**, 094604.

Pirozzoli, S. & Wei, T. 2025 On pressure fluctuations in the near-wall region of turbulent flows. *J. Fluid Mech.* **1010**, A10.

Rao, K. N., Narasimha, R. & Narayanan, M. A. B. 1971 The 'bursting'phenomenon in a turbulent boundary layer. *J. Fluid Mech.* **48** (2), 339–352.

Schewe, G. 1983 On the structure and resolution of wall-pressure fluctuations associated with turbulent boundary-layer flow. *J. Fluid Mech.* **134**, 311–328.

Schlatter, P. & Örlü, R. 2010 Assessment of direct numerical simulation data of turbulent boundary layers. *J. Fluid Mech.* **659**, 116–126.

Sillero, J. A., Jiménez, J. & Moser, R. D. 2014 Two-point statistics for turbulent boundary layers and channels at reynolds numbers up to $\delta^+ \approx 2000$. *Phys. Fluids* **26** (10).

Snarski, S. R. & Lueptow, R. M. 1995 Wall pressure and coherent structures in a turbulent boundary layer on a cylinder in axial flow. *J. Fluid Mech.* **286**, 137–171.

Thomas, A. S. W. & Bull, M. K. 1983 On the role of wall-pressure fluctuations in deterministic motions in the turbulent boundary layer. *J. Fluid Mech.* **128**, 283–322.

Townsend, A. A. 1976 *The structure of turbulent shear flow*, 2nd edn. CUP.

Tsuji, Y., Fransson, J. H. M., Alfredsson, P. H. & Johansson, A. V. 2007 Pressure statistics and their scaling in high-Reynolds-number turbulent boundary layers. *J. Fluid Mech.* **585**, 1–40.

Tsuji, Y., Imayama, S., Schlatter, P., Alfredsson, P. H., Johansson, A. V., Marusic, I., Hutchins, N. & Monty, J. 2012 Pressure fluctuation in high-Reynolds-number turbulent boundary layer: results from experiments and DNS. *J. Turbul.* (13), N50.

Tsuji, Y., Marusic, I. & Johansson, A. V. 2016 Amplitude modulation of pressure in turbulent boundary layer. *Int. J. Heat Fluid Flow* **61**, 2–11.

Wei, T., Fife, P., Klewicki, J. & McMurtry, P. 2005 Properties of the mean momentum balance in turbulent boundary layer, pipe and channel flows. *J. Fluid Mech.* **522**, 303–327.

Willmarth, W. W. 1975 Pressure fluctuations beneath turbulent boundary layers. *Annu. Rev. Fluid Mech.* **7**, 13–36.

Willmarth, W. W. & Wooldridge, C. E. 1962 Measurements of the fluctuating pressure at the wall beneath a thick turbulent boundary layer. *J. Fluid Mech.* **14**, 187–210.

Yu, M., Ceci, A. & Pirozzoli, S. 2022 Reynolds number effects and outer similarity of pressure fluctuations in turbulent pipe flow. *Int. J. Heat Fluid Flow* **96**, 108998.